\RequirePackage{fix-cm}
\documentclass[smallextended]{svjour3}       % onecolumn (second format)
\smartqed  % flush right qed marks, e.g. at end of proof
\usepackage{graphicx}
\usepackage{subfig}
\usepackage{natbib}
\usepackage{amsmath}
\usepackage{xcolor}
\usepackage{amssymb}
\usepackage{ulem}
\begin{document}

%\articletype{Original Research Article}% Specify the article type or omit as appropriate

\title{
Spatial Bayesian Hierarchical Modelling with Integrated Nested Laplace Approximation
}

\titlerunning{Spatial Bayesian Hierarchical Modelling with INLA}

\author{Nicoletta D'Angelo\thanks{Corresponding author:   Nicoletta D'Angelo, \email{nicoletta.dangelo@unipa.it}}         \and
        Antonino Abbruzzo \and
        Giada Adelfio %etc.
}

%\author{
%\name{Nicoletta D'Angelo*\textsuperscript{a}, Antonino Abbruzzo\textsuperscript{a} and Giada Adelfio\textsuperscript{a,b} \thanks{Email: nicoletta.dangelo@unipa.it, antonino.abbruzzo@unipa.it, giada.adelfio@unipa.it}}
%\affil{\textsuperscript{a}Dipartimento di Scienze Economiche, Aziendali e Statistiche, Università degli Studi di Palermo, Palermo, Italy\\
%\textsuperscript{b}Istituto Nazionale di Geofisica e Vulcanologia (INGV)\\
%* corresponding author}}

\institute{
			Nicoletta D'Angelo$^{1}$,  Antonino Abbruzzo$^{1}$,  Giada Adelfio$^{1,2}$ \at 1- Dipartimento di Scienze Economiche, Aziendali e Statistiche,  Universit\`{a} degli Studi di Palermo, Palermo, Italy\\
			2- Istituto Nazionale di Geofisica e Vulcanologia (INGV), Palermo, Italy}           %  \\
%             \emph{Present address:} of F. Author  %  if needed

\date{Received: date / Accepted: date}

\maketitle

\begin{abstract}
We consider latent Gaussian fields for modelling spatial dependence in the context of both spatial point patterns and areal data, providing two different applications. 
The inhomogeneous Log-Gaussian Cox Process model is specified to describe a seismic sequence occurred in Greece, resorting to the Stochastic Partial Differential Equations. The Besag-York-Mollié model is fitted for disease mapping of the Covid-19 infection in the  North of Italy. These models both belong to the class of Bayesian hierarchical models with latent Gaussian fields whose posterior is not available in closed form. Therefore, the inference is performed with the Integrated Nested Laplace Approximation, which provides accurate and relatively fast analytical approximations to the posterior quantities of interest.  
\end{abstract}

\section{Introduction}
\label{intro}
The fast-growing availability of vast sets of georeferenced data has generated a keen interest in suitable statistical modelling approaches to handle large and complex data. 
Bayesian hierarchical models have become a vital tool for capturing and explaining complex stochastic structures in spatial or spatio-temporal processes. Many of these models are based on latent Gaussian models, a vast and ﬂexible class ranging from generalized linear mixed to spatial and spatio-temporal models. 
Generally, closed-form expressions for the posterior distributions are not available for these complex models, and Markov Chain Monte Carlo (MCMC) algorithms can be used for inference \citep{hastings1970monte}, although this approach can be computationally demanding. Alternatively, the Integrated Nested Laplace Approximation (INLA)  has been developed as a computationally efficient algorithm for making inference on latent Gaussian models \citep{rue2009approximate}. Combining analytical approximations and numerical integration, INLA overcomes the convergence issues of the MCMC methods, providing faster computation. However, the use of analytical approximations may introduce errors in the computation of the posterior probabilities.
INLA solves inferential issues in many case studies with space-time applications. For example, it is applied to global climate data \cite{lindgren2011explicit},  in epidemiology \citep{Bisanzio2011}, in disease mapping and spread \citep{Schrodle2011, Schrodle2012}, to air pollution risk mapping \citep{cameletti2013spatio}, and in econometrics \citep{Gomez-Rubio2015a}. 
Recent review papers include \cite{lombardo2019numerical,cameletti2019bayesian,van2019new} and the recent book of \cite{gomez2020bayesian}.
More generally, INLA  is successfully applied to generalized linear mixed models \citep{Fong2010}, log-Gaussian Cox processes \citep{illian2012toolbox, gomez2015analysis} and survival models \citep{Martino2011}, amongst many other fields of application.
The recent monograph of Blangiardo and Cameletti (2015) reviews INLA in detail and gives many practical examples \citep{Blangiardo2015}.
Other available references for INLA are \cite{moraga2019geospatial,krainski2018advanced,wang2018bayesian,zuur2017spatial}.

The aim of this paper is to provide a review of the most common spatial models for epidemic phenomena, by means of well established standard methods. We explore the strength of the INLA approach in modelling spatial data available at different levels of detail, i.e. individual and aggregated.  According to the nature of data, different complex models are proposed in the literature.
In this paper, we focus on the latent Gaussian models applied to a seismic sequence occurred in Greece from 2005 and 2014, and the disease mapping of the Covid-19 infectious in the North of Italy from February to the end of April 2020.

%% Application to seimis seuence
The first application refers to a seismic sequence  described by fitting an inhomogeneous Log-Guassian Cox Process (LGCP) model.
The spatial seismic process belongs to the class of point pattern data, where the location of events in space are the observations of interest.
The aim is typically to learn about the mechanism that generates these events \citep{moller:98,diggle2013statistical,illian2008statistical,bakka2018spatial}. 
Different classes of point process models have been discussed in the literature,  from the simplest homogeneous Poisson process, which is used to describe uniform spatial randomness, to more complex models that generate aggregated or patterns exhibiting repulsion among points \citep{van2000markov,moller:03,illian2008statistical}.
The seismological process is a very complex phenomenon, characterized by fractal and multi-scale structures: its description needs complex models, able to account for environmental heterogeneity and detecting interactions at several spatial scales, usually in terms of clustering. Indeed, the distribution of seismic events is relatively complex, and different sources of earthquakes (faults, active tectonic plate and volcanoes) may produce events with different spatial displacements and orientations. The multi-scale interaction between the earthquakes and the relationship between the conditional intensity and the geological information available in the study area can be a crucial point of research \citep{siino:adelfio:mateu:16}. When dealing with point-reference data a computationally effective estimation approach is based on the Stochastic Partial Differential Equation (SPDE) method \citep{lindgren2011explicit}. This method consists of representing a continuous spatial process, e.g. a Gaussian Random Field with the Matérn covariance function, as a discretely indexed spatial random process, e.g. a Gaussian Markov Random Field, providing substantial computational advantages. Therefore, in this paper, we explore the advantages the LGCP fitting to the considered Greek complex seismic data, characterized by a strong clustered structure,  %related also to external covariates, 
using the SPDE models, as in \cite{simpson2016going}.

%areal data
In the second application, the mapping of the Covid-19 infectious disease in the North of Italy, caused by the most recently discovered coronavirus, is provided by fitting the BYM model, 
thought the INLA estimation approach. This new virus and the triggered disease were unknown before the outbreak began in Wuhan, China, in December 2019. However,  Covid-19 is still a tragic pandemic spreading out around the world very quickly. Italy is one of the most affected countries.
Recent papers, dealing with the outbreak and spread of Covid-19 infectious disease, concern analyses both at country level \citep{kang2020spatial} and global scale \citep{liao2020tempo}. These papers mostly provide descriptive analysis in terms of the geographic distribution of the Covid-19 cases. Few attempts have been made in proposing modelling approaches taking into account the spatial component of the phenomenon. For instance, in \cite{ramirez2020spatial} the authors study the epidemic spread in Iran through spatial linear models in order to identify the variables that have significantly impacted the number of cases of Covid-19. 
In this paper, we analyse the Italian data on Covid-19, which are collected at an aggregate level and comprise the daily counts of the infected people in the regions and provinces. This data is a typical example of areal data which can be modelled by the Besag–York–Mollié (BYM) model \cite{besag1991bayesian}.
This is widely used in disease mapping based on areal data, for its ability to recover risk surfaces and identify areas of high-risk areas or hot-spots, starting from aggregated data.
Therefore, in this paper, the BYM model is employed for accounting for the neighbourhood structure of the available count data, modelling the number of cases per district and identifying the high-risk areas, proving that relevant information can still be elicited dealing with  aggregated data, without external covariates. Furthermore, in the proposed analysis, we explore whether there is a difference between the period proceeding the lock-down and the subsequent one in terms of the risk disease.

%structure
The paper is structured as follows. A brief overview of the INLA approach is given in Section \ref{section:INLA}. In Section \ref{section:models}, inference on the spatial LGCP, combining the INLA and SPDE approaches and on the Besag-York-Mollié with INLA, are reported. The applications to earthquake data and disease mapping are provided in Section \ref{section:application}. Finally, the paper ends with a discussion in Section \ref{section:conclusions}.

\section{Background on the Integrated Nested Laplace Approximation}
\label{section:INLA}
INLA is designed for making inference of \textit{Latent Gaussian Models}, a very wide and ﬂexible class of models, ranging from generalized linear mixed to spatial and spatio-temporal models. 
Let $Y_i, i = 1, \ldots, n$, be the response variable and assume that its distribution belongs to the exponential family, in which the linear predictor $\eta_i$ can include terms of covariates in an additive way, 
\begin{equation}
\label{eq:addlin}
\eta_i=\alpha+\sum_{k=1}^{n_{\beta}}\beta_kz_{ki}+\sum_{j=1}^{n_f}f^{(j)}(w_{ji}),
\end{equation}
where $g(\cdot)$ is the link function, $\eta_i = g(\mu_i) = g(E[Y_i])$, $\alpha$ is the intercept, $\boldsymbol{\beta} = (\beta_1, \ldots, \beta_{n_\beta})$ are the regression parameters of covariates $z = (z_1, \ldots, z_{n_\beta})$, and $f=\left\{f^{(1)}(\cdot),f^{(2)}(\cdot),\dots,f^{(n_f)}(\cdot)\right\}$ is a set of $n_f$ functions of some covariates $w$ that may  be  related to $\eta$ also by a non-linear relationship.
This non-linear relation can be accounted for by several statistical approaches, such as random (\textit{iid}) effects, spatially or temporally correlated effects, smoothing splines, by varying the functional form of the  $f(\cdot)$s. In other words, this formulation can be used for a wide range of models, e.g. standard and hierarchical regressions,  spatial and spatio-temporal models, etc.
Let $\boldsymbol{\theta} = (\theta_1, \ldots, \theta_{1+n_\beta+n_f}) = (\alpha, \boldsymbol{\beta}, f)$ denote the vector of the model parameters, called latent field, and assume that the multivariate  distribution belongs to a Gaussian Markov Random Field (GMRF), with zero mean and with a sparse precision matrix $Q(\boldsymbol{\psi})$, where $\boldsymbol{\psi}$ denotes a vector of hyperparameters, which are not necessarily Gaussian. 
The main goal of the INLA-methodology is to obtain the marginal distributions for the elements of the latent field $p(\boldsymbol{\theta}|\textbf{y})$ and the hyperparameters $p(\boldsymbol{\psi}|\textbf{y})$, and use these to compute posterior summary statistics. This is achieved by exploiting the computational properties of the GMRF and the Laplace approximation for multidimensional integration, assuming the conditional independence of the observed variables \textbf{y} given the latent field  $\boldsymbol{\theta}$ and the hyperparameters  $\boldsymbol{\psi}$.
For further details on the inferential process we refer to \cite{rue2009approximate,martins2013bayesian,blangiardo2013spatial}.
%A toy example of INLA for a multiple linear regression model is reported in the Appendix \ref{section:app}.

\section{Spatial models with INLA}
\label{section:models}
Spatial models can be distinguished into continuous and discrete domain models.  When geocodes are available,  Log-Gaussian Cox Processes (LGCPs), that are continuous domain models, can be used \citep{konstantinoudis2020discrete}.
When individual data are not available, models for areal data can be applied. In particular, the Besag–York–Mollié (BYM) model is a discrete domain model, used for describing count data per spatial unit.
Both models are widely considered in disease mapping, as calculating and visualising disease risk across space is an important exploratory tool in epidemiology. Indeed, the BYM model can be seen as a special case of the LGCP, where the relative risk is assumed to be piecewise constant within regions.
\cite{li2012log} presents a methodology for modelling aggregated disease incidence data with the spatially continuous LGCP, using a data augmentation step to sample from the posterior distribution of the exact locations at each step of an MCMC algorithm, and modelling the exact locations with a LGCP.

\subsection{Continuous domain models for spatial point processes}
\label{section:continuous}

A spatial point pattern $N$ is an unordered set $\textbf{x}=\{\textbf{x}_1,\dots,\textbf{x}_n\}$  of points $\textbf{x}_i$ where $n(\textbf{x})=n$ denotes the number of points, not fixed in advance.
If $\textbf{x}$ is a point pattern and $D \subset \mathbb{R}^d $  is a region such that $|W|< \infty$ and $d=2$, we write $ \textbf{x} \cap W$ for the subset of $\textbf{x}$ consisting of points
that fall in $W$ and $n(\textbf{x} \cap W)$ for denoting the number of points of $\textbf{x}$ falling in $W$.
A point process model assumes that $\textbf{x}$ is a realization of a finite point process $N$ in $W$ without multiple points. 
 A point location in the plane is denoted by a lower case letter like $\textbf{s}$. Any location $\textbf{s}$ can
be specified by its Cartesian coordinates $\textbf{s} = (s_1 ,s_2 )$. 
The coordinates of the observed events depend on an underlying generating spatial process, which is often characterized by its intensity function $\lambda(\textbf{s})$. 
The intensity function measures the average number of events per unit of space, and it could also depend on covariates and other effects.
Earthquakes are a typical example of point pattern data in which one is interested in studying the spatial intensity of the seismic activity observed in a given area, and the LGCP is  largely used for analysing spatial point pattern for earthquake data  \citep{moller:98, siino2018joint}. 

The LGCP belongs to the class of the Cox processes which are models for point phenomena that are environmentally driven and have a clustered structure.
They are Poisson processes with a random intensity function depending on unobservable external factors. 
The point process $N$ is said to be a Cox process driven by $\Lambda$, if the conditional distribution of the point process $N$ given a realisation $\Lambda(\textbf{s})=\lambda(\textbf{s})$ is a Poisson process on $W $ with intensity function $\lambda(\textbf{s})$. 
Therefore, given an observed point pattern $\textbf{y}$, the LGCP model is defined as
\begin{align}
\textbf{y}|\lambda(\textbf{s}) \sim e^{|W|-\Lambda}\prod_{\textbf{s}_i \in \textbf{y}} \lambda(\textbf{s})\nonumber \\
        \log(\lambda(\textbf{s}))=X(\textbf{s})'\boldsymbol{\beta}+U(\textbf{s})\nonumber\\
U(\textbf{s})\sim GRF(0,\Sigma(\sigma^2,r)).
\label{eq:LGCP}
\end{align}
Let $\textbf{s}$ be any location in a study area $W$ and let $U(\textbf{s})$ be the random spatial effect at that location. $U(\textbf{s})$ is a stochastic process, with $\textbf{s} \in W$, where $W \in \mathbb{R}^2$.  Since $U(\textbf{s})$ is assumed to be continuous over space, then it is a continuously-indexed Gaussian Random Field (GRF). This implies that it is possible to collect these data at any finite set of locations within the study region. We  denote by $u(s_i), i=1,\dots,n$ a realization of $U(\textbf{s})$ at $n$ locations, which is assumed to have a multivariate Gaussian distribution. Therefore, for the specification of the model, it is necessary to define its mean and covariance. 

 Since a zero mean process is usually assumed, its multivariate distribution is completely determined by the spatial covariance function.
The range parameter $r$ of the spatial covariance function controls the roughness of the relative risk surface, i.e. larger values provide  stronger correlation and smoother surfaces, while $\sigma^2$ is the marginal variance of the process.
The LGCP model allows  to account for further additive  external covariates. In the context of spatial point processes, the $X(\textbf{s})$ are referred to as spatial covariates, meaning that their value is assumed to be observable, at least in principle, at each location $\textbf{s}$ in the window $W$. 
For inferential purposes, their values must be known at each point of the data point pattern and at least at some other locations.
In this LGCP model specification, the latent field is represented by $\boldsymbol{\theta} = (\boldsymbol{\beta},\boldsymbol{u})$ and the hyperparameters are $\boldsymbol{\psi} = (\sigma^2,r)$.

In general, the Cox model is estimated by a two-step procedure, involving first the intensity and then the cluster or correlation parameters. In the first step, a Poisson model with the same model formula is fitted to the point pattern data, providing the estimates of the coefficients of all the terms in the model formula that characterize the intensity. Then, in the second step, the estimated  intensity is  taken as the true one and the cluster  or correlation parameters are estimated by the method of minimum contrast \citep{pfanzagl1969measurability,eguchi1983second,diggle1979parameter,diggle1984monte,siino2018joint}, or the Palm likelihood \citep{ogata1991maximum,tanaka2008parameter} or composite likelihood \citep{guan2006composite}.

Unlike traditional inferential methods for the LGCP models, the SPDE approach  uses the exact locations of the point pattern  and provides a continuous approximation of the latent field.
In the case of geostatistical data,  $U(\textbf{s})$ has a multivariate Normal distribution with zero mean and spatially structured covariance matrix $\Sigma$, whose generic elements is $\Sigma_{ij}=\sigma^2Cov(U(s_i),U(s_j))$ where 
\begin{equation}
\label{eq:covmat}
Cov(U(s_i),U(s_j))=\frac{1}{\Gamma(\nu)2^{\nu-1}}(\kappa\Delta_{ij})^{\nu}K_{\nu}(\kappa\Delta_{ij})
\end{equation}
is the isotropic Matérn spatial covariance function \citep{cressie1992statistics}, depending on the Euclidean distance between the locations $\Delta_{ij}=\|s_i-s_j\|$. The parameter $K_{\nu}$ denotes the modified Bessel function of second kind and order $\nu>0$, which measures the degree of smoothness of the process and is usually kept fixed. Conversely, $\kappa>0$ is a scaling parameter related to the range $r$, i.e. the distance at which the spatial correlation becomes almost null. Typically, the empirically derived definition $r=\frac{\sqrt{8\nu}}{\kappa}$ is used \citep{lindgren2011explicit}, with $r$ corresponding to the distance at which the spatial correlation is close to 0.1 for each $\nu$. 
Other models can be considered for the spatial covariance function, but in this paper we only focus on the Matérn model since it is used in the SPDE approach proposed by \cite{lindgren2011explicit}. However, this point does not represent a limitation, since, as stated by \cite{guttorp2006studies}, the Matérn family is a very flexible class of covariance functions able to cover a wide range of spatial fields.
In detail, the SPDE is a  computationally effective tool  dealing with point-reference data. It consists in representing a continuous spatial process, e.g. the GRF $U(\textbf{s})$ with the Matérn covariance function defined in Equation (\ref{eq:covmat}), as a discretely indexed spatial random process (e.g. a GMRF) by  a basis function representation defined on a triangulation of the domain $W$, that in turn, produces substantial computational advantages.
We refer to \cite{lindgren2011explicit} for a complete description.
The \texttt{inlabru} package \citep{bachl2019inlabru} provides a simple interface for the analysis of point patterns with INLA and SPDE models. However, in the next sections we provide  full details to fit the model according to \cite{simpson2016going}.

\subsection{Discrete domain models for areal data}
\label{section:discrete}

The Besag-York-Mollié model \cite{besag1991bayesian} is an extension of the ICAR (Intrinsic Conditional Autoregressive) model, obtained by adding a spatially unstructured random effect to the already given spatially structured random effect. The latter is a realisation of a GMRF with zero mean and a sparse precision matrix capturing strong spatial dependence. The unstructured random effect may be seen as a collection of independent random intercepts for the various areal units. This specification leads to a piecewise constant risk surface which depends on the spatial unit selected and assumes uniform risk across this spatial unit.
For some discussions of the scaling issues, refer to \cite{freni2018note,riebler2016intuitive}.

Let $Y_i$ be the random variable number of cases in the region $D_i, i=1,\dots,n$,  and $E_i$ the corresponding
expected cases count for the $i$-th
spatial unit. The BYM model in its general form is 
\begin{align}
    Y_i   \sim Poisson(E_i\lambda_i)\nonumber \\
        \log(\lambda_i)  = \alpha+X_i'\boldsymbol{\beta}+U_i\nonumber \\
        U_i  \sim BYM(\sigma^2_v,\sigma^2_{\nu}),
        \label{eq:BYM}
\end{align}
where the linear predictor is specified on the logarithmic scale. As for the LGCP model specification, also the BYM accounts for further additive external covariates as potential risk factors.
Here $X_i$ are region level covariates (e.g. average income and education levels) with coefficients $\boldsymbol{\beta}$, and $\alpha$ is the intercept quantifying the average of the  cases in all the $n$ regions.

The random spatial process $U_i$ is the sum of an independent
Gaussian process $\nu_i$ with variance $\sigma^2_\nu$ and a GMRF $v_i$ with
variance $\sigma^2_v$. For each region, the value of the GMRF component depends on the average from
the neighboring regions 
  \begin{equation}
    v_i|v_{-i},\sigma^2_v\sim GMRF\Biggl(\frac{\sum_{j \in d_i}v_j}{d_i},\frac{\sigma^2_v}{d_i}\Biggl),
    \label{eq:str}
\end{equation}
where $d_i$ is the number of areas which share boundaries with the $i$-th one. Two regions are usually defined as neighbours if they share a common border.
Using the notation of the Equation (\ref{eq:addlin}),
$v_i=f^{(1)}(i)$ and $\nu_i=f^{(2)}(i)$ are two area-specific effects.

More in detail, the parameter $\nu_i$ represents the unstructured residual, modelled as
\begin{equation}
    \nu_i|\sigma^2_{\nu}\sim N(0,\sigma^2_{\nu}).
    \label{eq:unstr}
\end{equation}
When  the ratio  $\sigma^2_{\nu}/\sigma^2_v$ increases, the $U_i$'s spatial dependence increases as well, providing a smoother surface.
The latent field is represented by $\boldsymbol{\theta} = (\alpha,\beta,\boldsymbol{u})$ and the hyperparameters are $\boldsymbol{\psi} = (\sigma^2_v,\sigma^2_{\nu})$.

\section{Spatial models estimation for two real datasets}\label{section:application}

In this section, we report the study of two spatial datasets, referring both to the continuous and discrete domain models, according to the nature of data.
All the analyses are carried out using the \texttt{R-INLA} package \citep{rue2009approximate,martins2013bayesian,Blangiardo2015}  of the software R \citep{R}. In particular, for the  inhomogeneous Log-Gaussian Cox Process fitting tools with the SPDE, we refer to \cite[Chapter~4]{krainski2018advanced}. For the specification of the Besag-York-Mollié model with INLA, we refer to \cite[Chapter~6]{Blangiardo2015}.  All the codes of the  carried out analyses throughout the paper are available on request.

\subsection{Greek Seismicity characterization by a spatial LGCP}

\begin{figure}
	\centering
	\includegraphics[width=\textwidth]{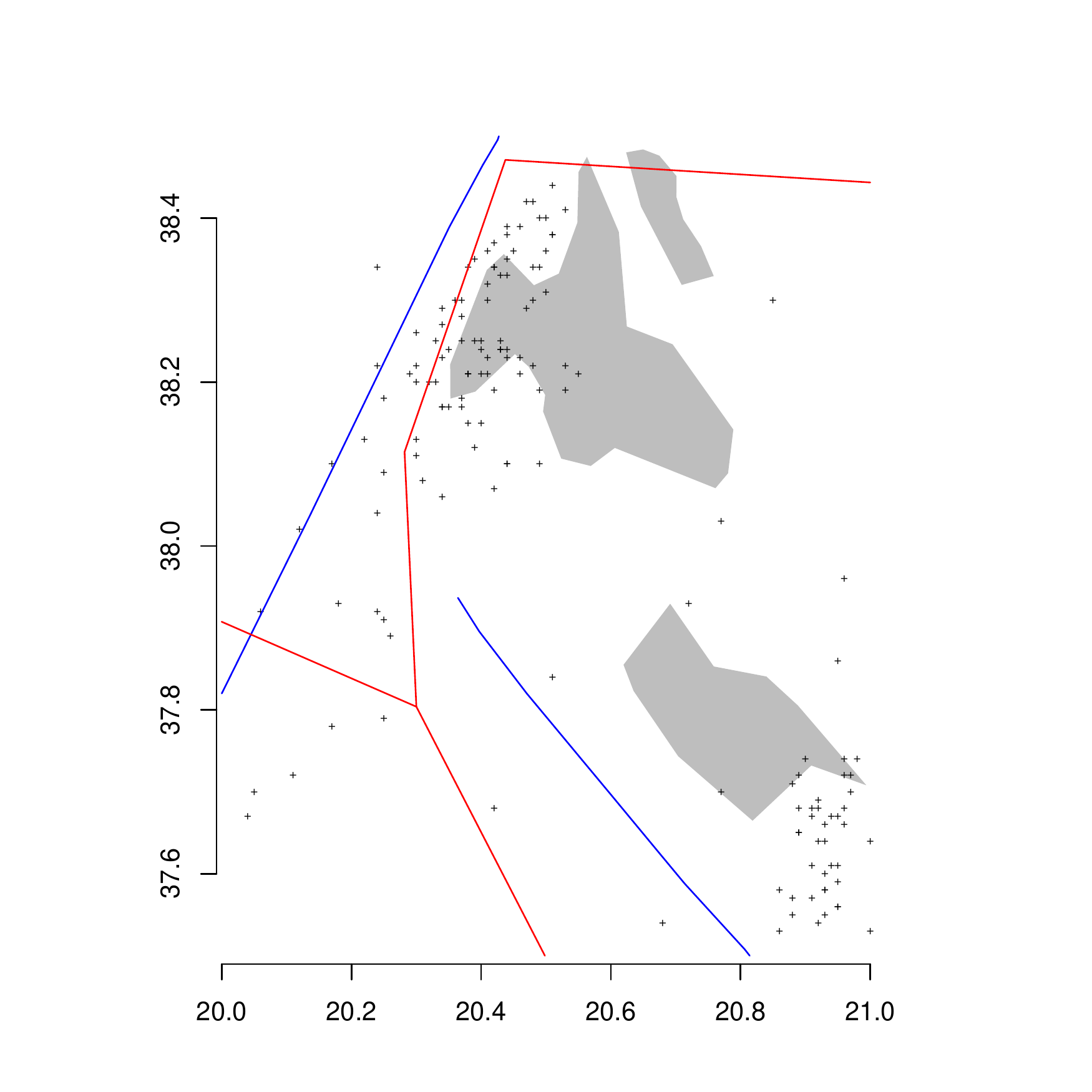}
	\caption{Earthquakes occurred in Ithaki, Kefalonia and Zakynthos between 2005 and 2014. The plate boundary is displayed in red while the faults are displayed in blue.}
	\label{fig:gg3}
\end{figure}

\begin{figure}
	\centering
	
	\subfloat[$D_{f}$]{\includegraphics[width=0.33333\textwidth]{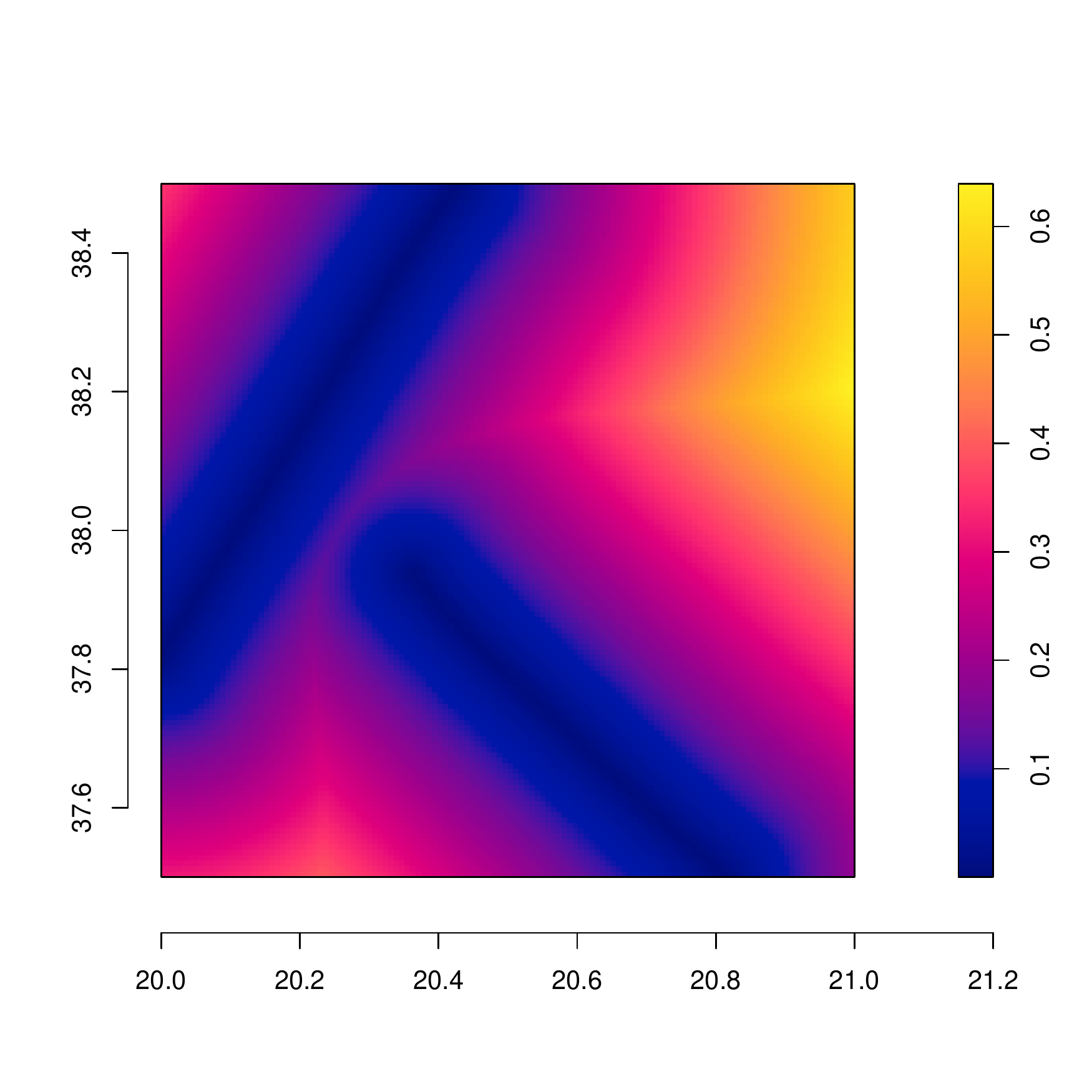}}
	\subfloat[$D_{pb}$]{\includegraphics[width=0.33333\textwidth]{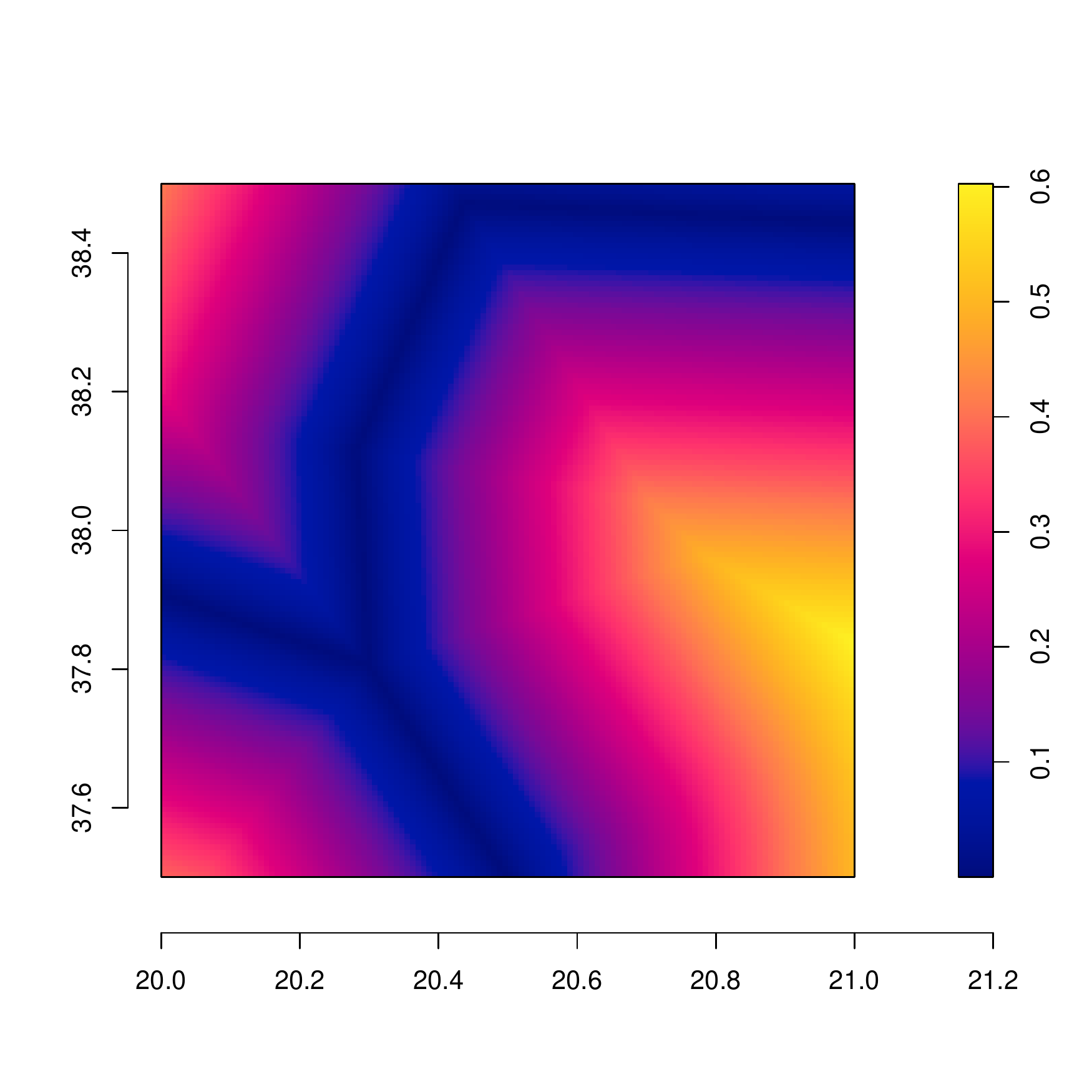}}
	\subfloat[$D$]{\includegraphics[width=0.33333\textwidth]{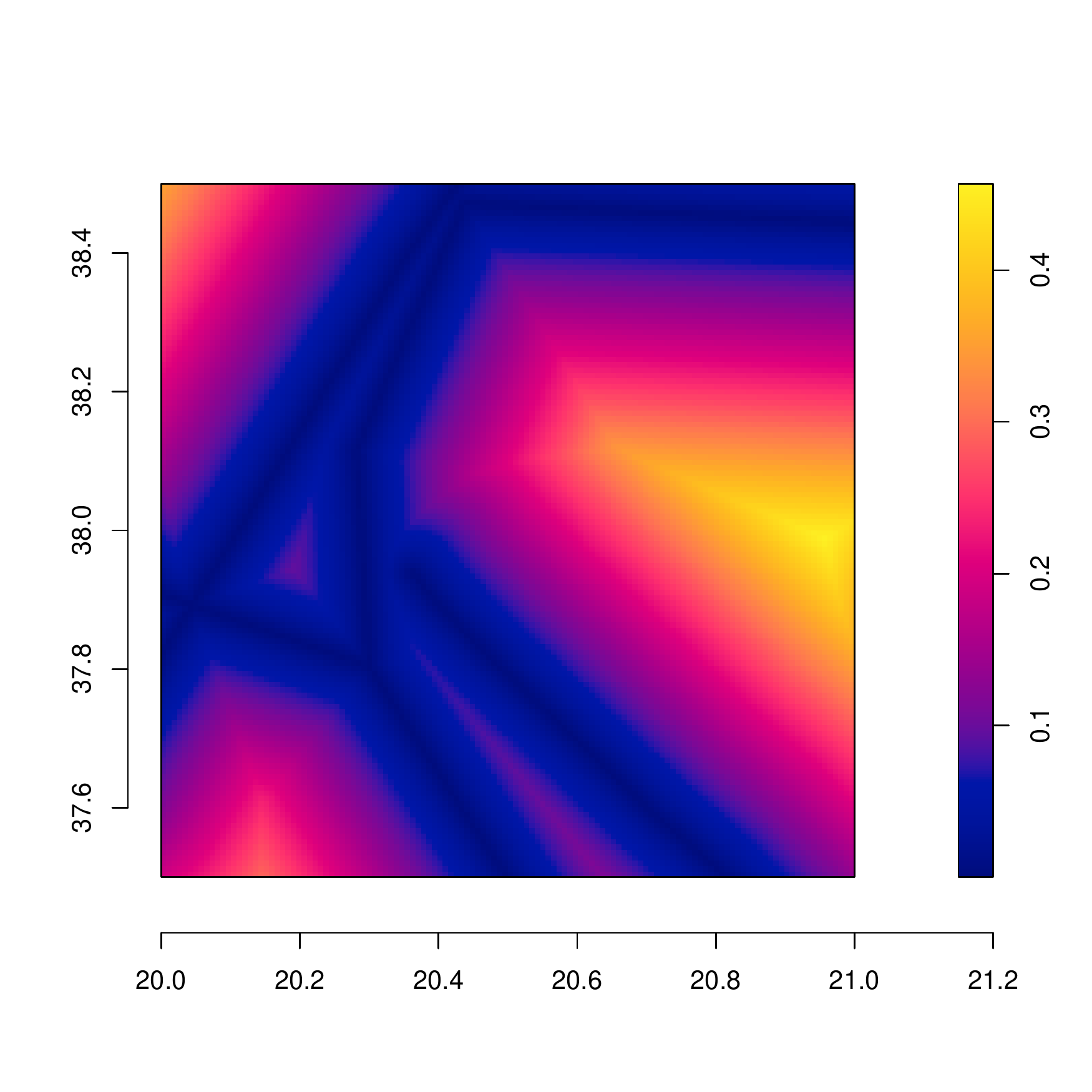}}
	\caption{Spatial covariate's surfaces}
	\label{fig:covs}
\end{figure}

The original analysed data concern 1105 earthquakes occurred in Greece  between 2005 and 2014. Only seismic events with a magnitude larger than 4 are considered in this study, and  the analysis in this paper are marginal with respect to the  time, focusing on the spatial dependence of events.
Here we analyse a small spatial window, focusing on  events with the most inhomogeneous behaviour. In detail, the study area is situated in the Ionian Sea, and it comprehends three of `the Seven Islands', namely Ithaki, Kefalonia and Zakynthos, from North to South. 
In Figure \ref{fig:gg3}, the 149 analysed earthquakes, together with plate boundary  (in red)  and faults  (in blue), are displayed.
In this area, earthquakes appear to be clustered in the Western area of Kefalonia island and South to the Zakynthos island.
Since the distance from the seismic sources is expected to influence the intensity of earthquakes, the available spatial covariates are taken into account. These are the Distance from the faults ($D_f$) and the Distance from the plate boundary ($D_{pb}$), 
computed as the shortest Euclidean distances form the spatial location $\textbf{s}$ of  events and the   map of geological information \citep{baddeley2015spatial}. More in detail, the distance function of a set of points $N$ is the mathematical function $f$ such that, for any two-dimensional spatial location $\textbf{s}$ is the function value $f(\textbf{s})$ is the shortest distance from $\textbf{s}$ to $N$. In particular, only the information about the top segment of the fault is used to compute the distances. Finally, also the spatial covariate $D_{ns}$,  Distance from the nearest seismic source, is computed and included in the analysis. %, obtained after the superimposition of the line segment patterns of the two available seismic sources. %before computing the Euclidean distance. 
%wondering  if the effect of the distance on the intensity of the process varies according to the two seismic sources.
 The spatial covariate's surfaces are displayed in Figure \ref{fig:covs}, where a  brighter  colour represents a larger  distance from the seismic source.

%\subsection{Model selection}

First, we fit a spatial LGCP model, without additional covariates in the linear predictor, as follows:
\begin{equation} 
\log \lambda(s_i) = \beta_0+u(\textbf{s}_i).
\label{eq:eq5}
\end{equation}

\begin{figure}
	\centering
	\includegraphics[width=\textwidth]{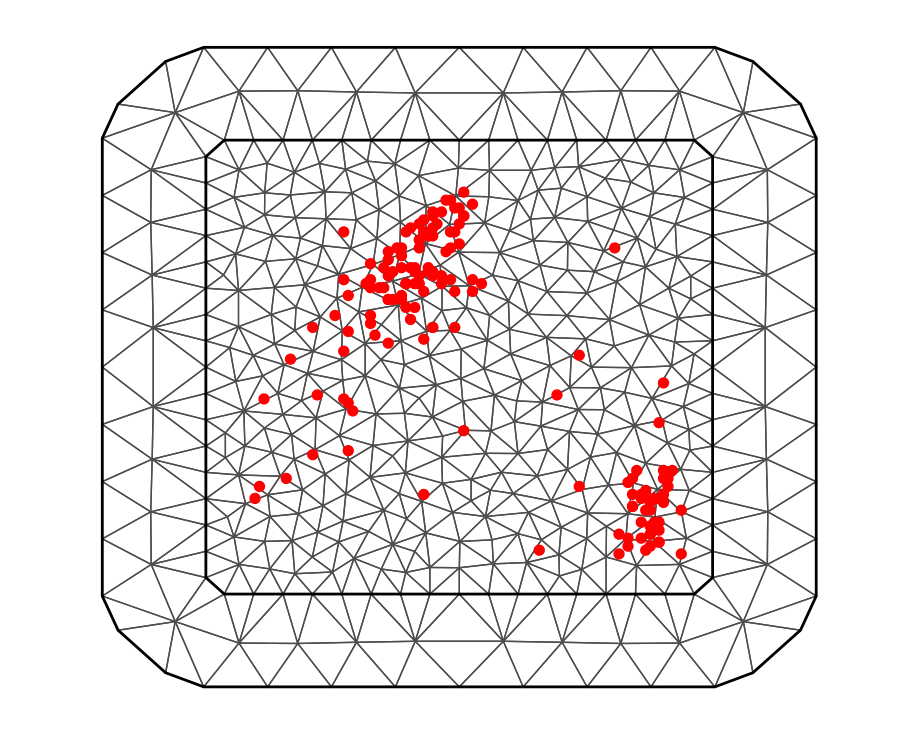}
	\caption{Mesh used to fit the Log-Gaussian Cox Process model to the  analysed point pattern}
	\label{fig:mesh3}
\end{figure}

The SPDE approach for point pattern analysis defines the model at the nodes of the mesh, so this is built on the entire domain extent, with the largest allowed triangle edge length equal to 0.1 for the interior edges and 0.4 for the exterior (see Figure \ref{fig:mesh3}).
Penalized Complexity priors, derived as in \cite{fuglstad2019constructing}, are used for the range $r$ and for the standard deviation $\sigma^2$ of the Matérn covariance. In particular, the prior of the range $r$  is set by defining $(r_0,p_r)$ such that  $P(r < r_0)=p_r$, and the  the penalized complexity prior for the standard deviation $\sigma$ is defined by the pair $(\sigma_0,p_s)$, such that $P(\sigma > \sigma_0)=p_s$. As a prior assumption the probability that the range is lower than 0.05 is large (i.e., $r_0=0.05$ and $p_r=0.01$ ) and that the probability that $\sigma$ is higher than 1 is small (i.e., $\sigma_0=1$ and $p_s=0.01$).

\begin{figure}
	\centering
		\subfloat[$\beta_0$]{\includegraphics[width=0.33333\textwidth]{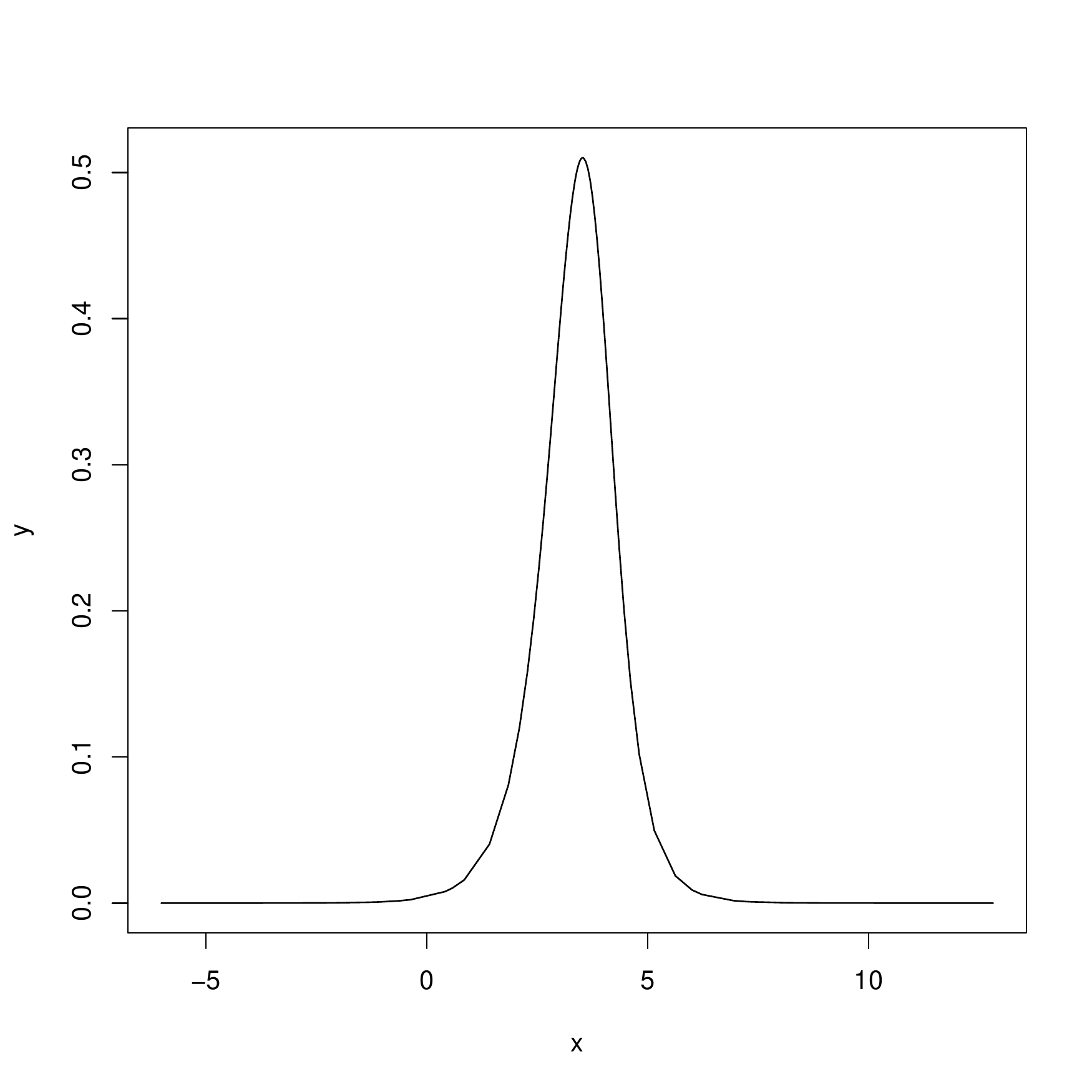}}
	\subfloat[$r$]{\includegraphics[width=0.33333\textwidth]{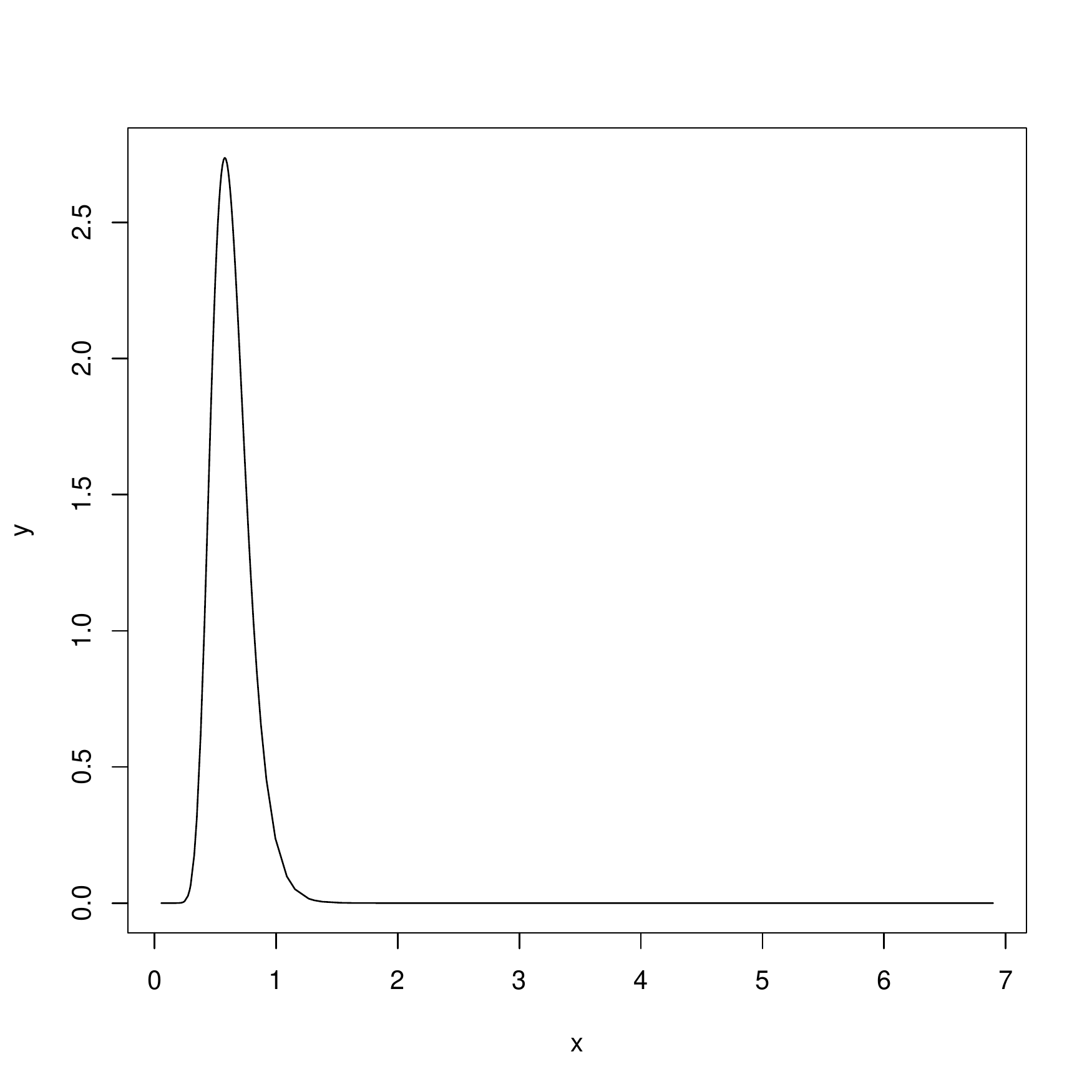}}
	\subfloat[$\sigma^2$]{\includegraphics[width=0.33333\textwidth]{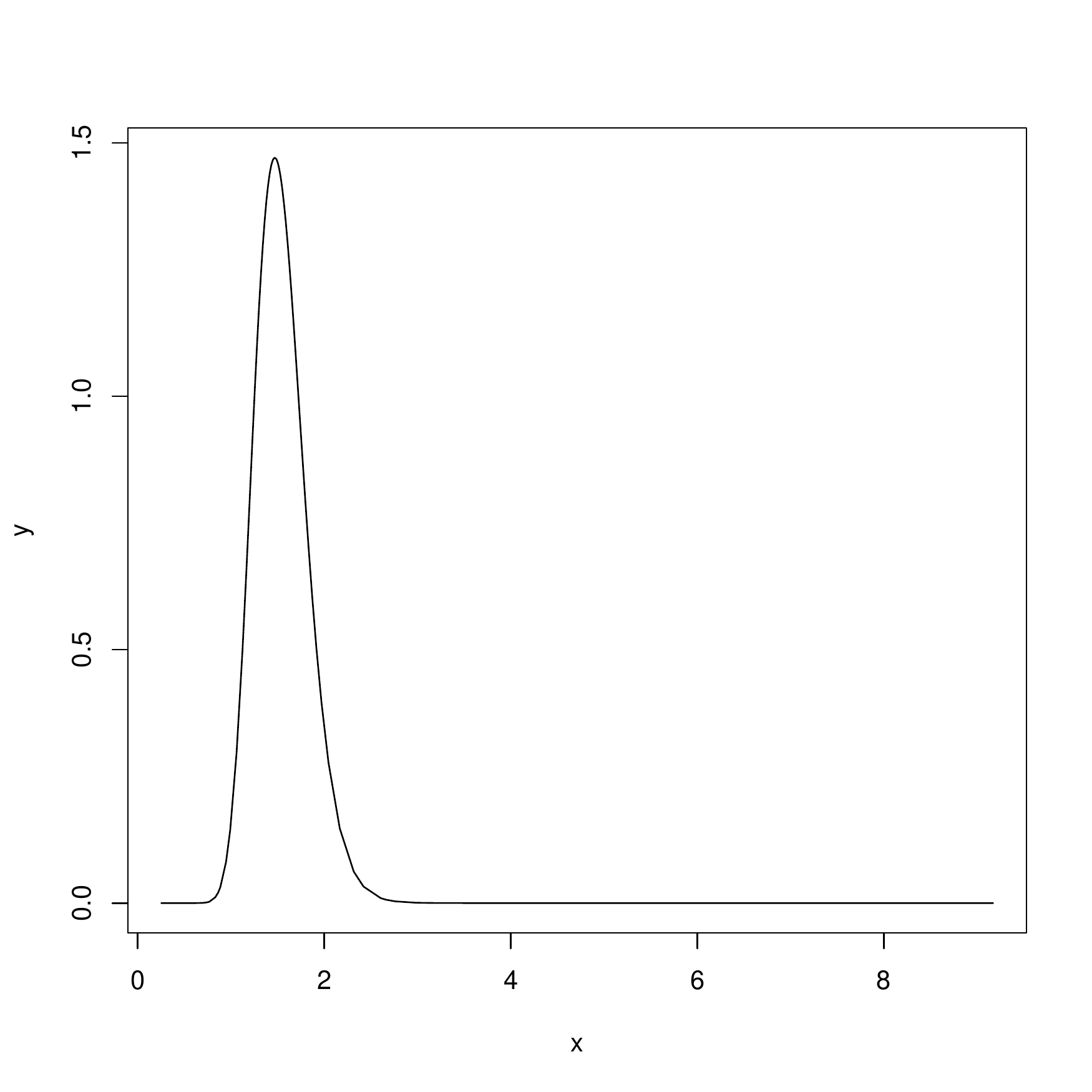}}
	\caption{Posterior distributions of the parameters and hyperparameters of the fitted LGCP model \ref{eq:eq5}}
	\label{fig:betas2}
\end{figure}

The posterior distributions of the parameters and hyperparameters of the model in Equation (\ref{eq:eq5}) are shown in Figure \ref{fig:betas2}. The summary statistics of the posterior distribution of the intercept $\beta_0$, that in this context represents the expected number of points in the unit area, are reported in Table \ref{tab:sumbetas}, together with the summary statistics of the posterior distributions of the hyperparameters $r$ and $\sigma^2$. 
%These results seem   reasonable and plausible.

\begin{table}
\caption{Summary statistics of the parameters and hyperparameters of the LGCP model \ref{eq:eq5}}
	\label{tab:sumbetas}
	\begin{tabular}{rrrrrrr}
		\hline
		& mean & sd & 0.025quant & 0.5quant & 0.975quant & mode  \\ 
		\hline
		$\beta_0$ & 3.40 & 0.92 & 1.42 & 3.45 & 5.15 & 3.53  \\
				\hline
		$r$ & 0.63 & 0.16 & 0.38 & 0.61 & 0.99 & 0.58 \\ 
		$\sigma^2$ & 1.54 & 0.28 & 1.06 & 1.52 & 2.17 & 1.47 \\ 
		\hline
	\end{tabular}

\end{table}

%On the basis of our experience, the parameter values obtained are reasonable and plausible.

Then, several more complex models are fitted, adding covariates to the linear predictor. These models differ only for the specification of the linear predictor. We consider 
\begin{equation}
\log \lambda(s_i) = \beta_0+\beta_1D_{ns}+u(s_i)
\label{eq:a}
\end{equation}
\begin{equation}
\log \lambda(s_i) = \beta_0+\beta_1D_{f}+\beta_2D_{pb}+u(s_i)
\label{eq:b}
\end{equation}
\begin{equation}
\log \lambda(s_i) = \beta_0+\beta_1D_{f}+\beta_2D_{pb}+\beta_3D_{f}D_{pb}+u(s_i)
\label{eq:selected}
\end{equation}

The DIC criterion \citep{spiegelhalter2002bayesian} and CPO values are used for comparing  the estimated models, as reported in Table \ref{tab:dic}. 
\begin{table}
	\caption{DIC and CPO values for the fitted LGCP models}
		\label{tab:dic}
	\begin{tabular}{rrrrr}
		\hline
		& \multicolumn{4}{c}{Models}\\
		\cline{2-5}
		& \ref{eq:eq5} & \ref{eq:a} & \ref{eq:b} & \ref{eq:selected} \\
		\hline
		DIC & -1509.89 & -1512.20 & -1514.50 & -1514.81 \\ 
		CPO & -671.85 & -649.26 & -615.90 & -595.40 \\ 
		\hline
	\end{tabular}

\end{table}    
On the one hand, the DIC is a generalization of the AIC and therefore the preferred models have low DIC values.
On the other hand, the CPO estimates the probability of observing $y_i$ in the future after having already observed $y_{-i}$, and therefore we chose the models with high CPO values.

\begin{figure}
	\centering

	\subfloat[$\beta_0$]{\includegraphics[width=0.33333\textwidth]{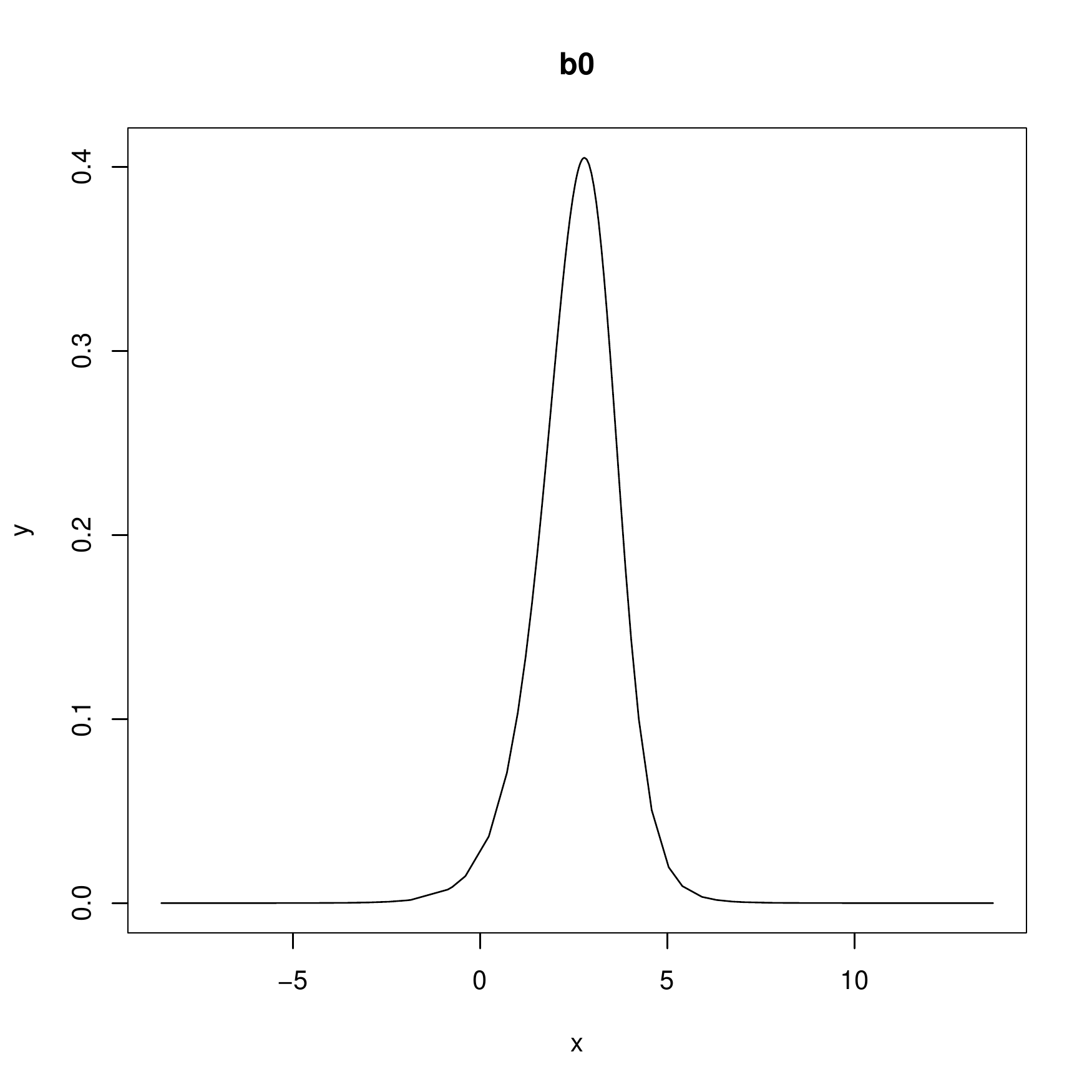}}
	\subfloat[$\beta_1$]{\includegraphics[width=0.33333\textwidth]{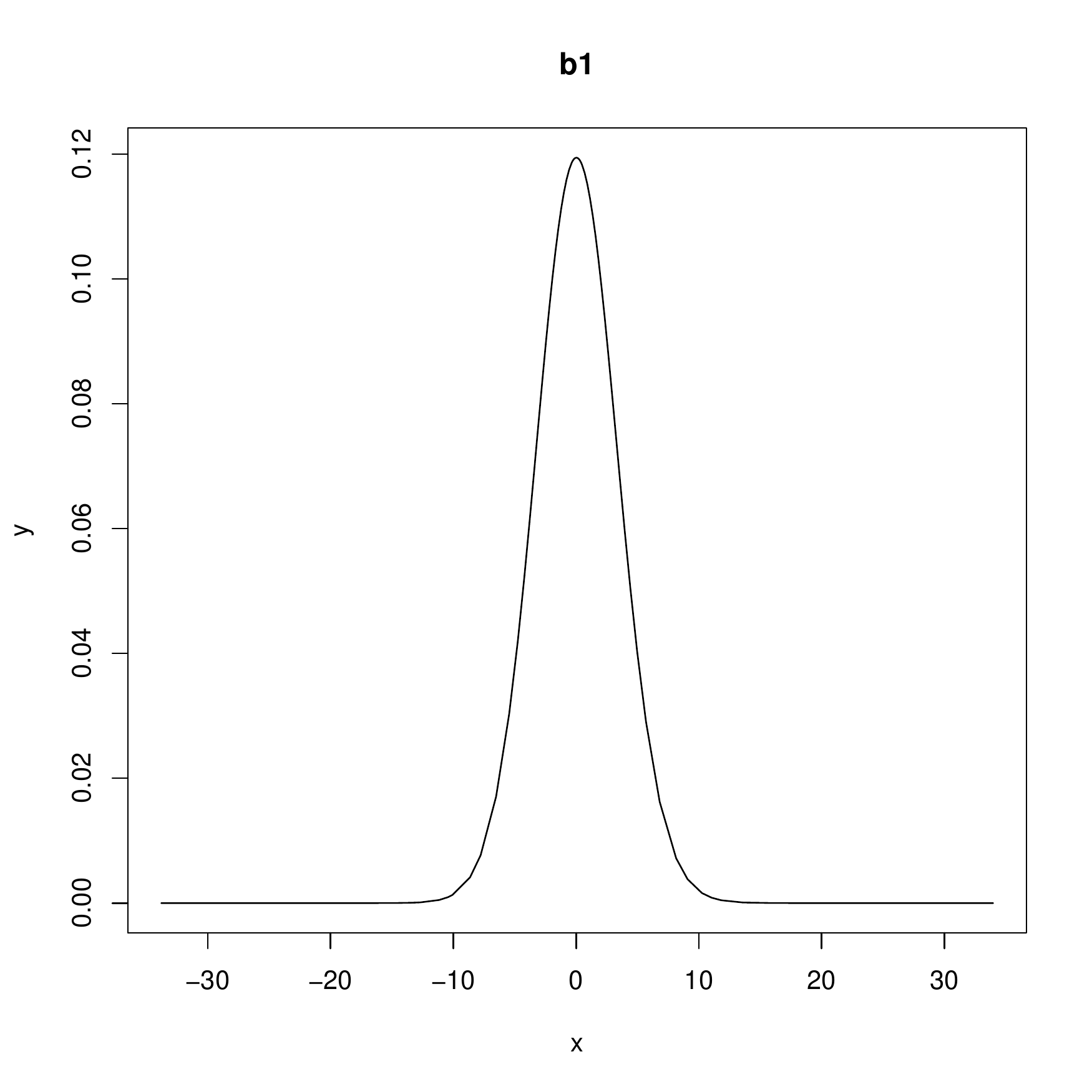}}        \subfloat[$\beta_2$]{\includegraphics[width=0.33333\textwidth]{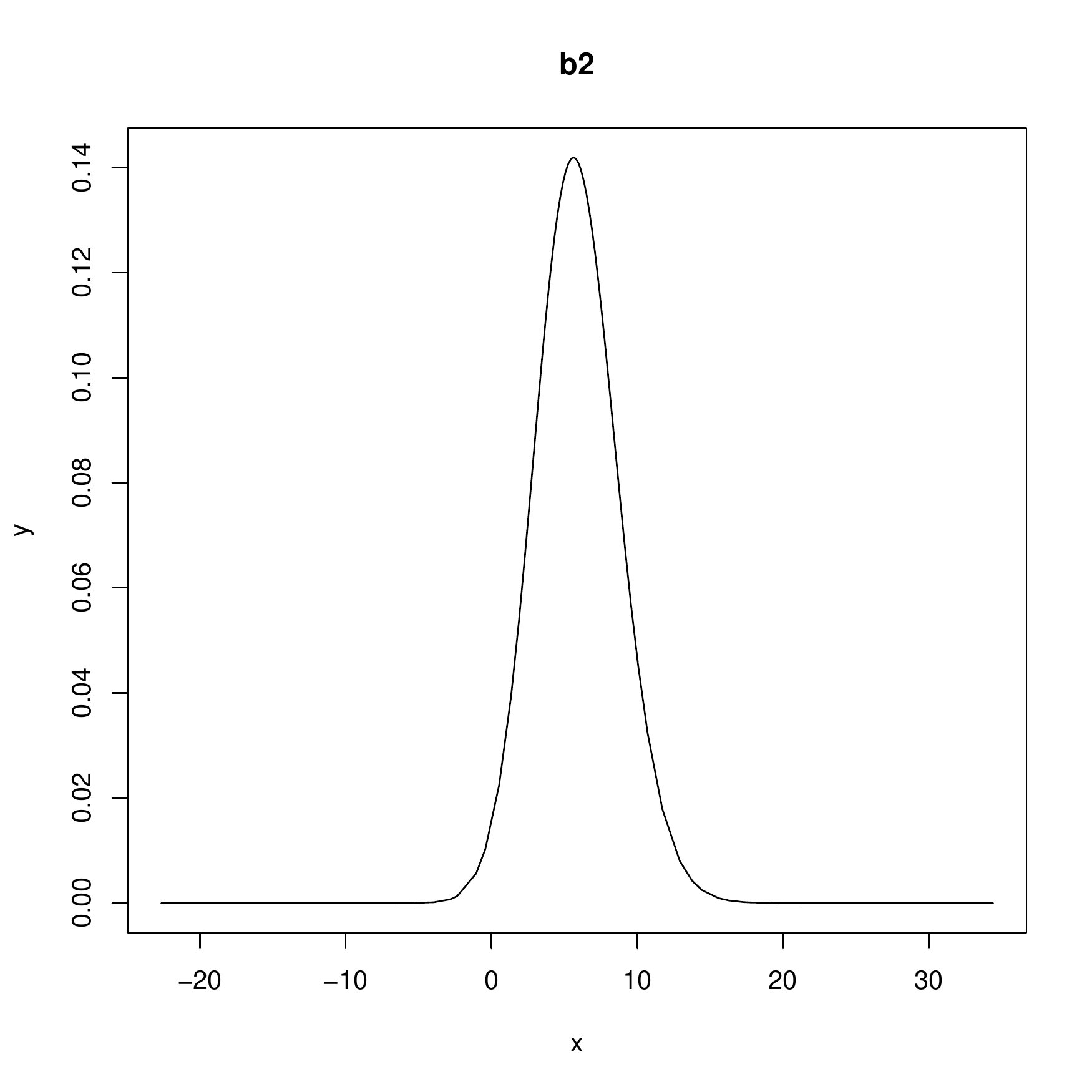}}\\
	\subfloat[$\beta_3$]{\includegraphics[width=0.33333\textwidth]{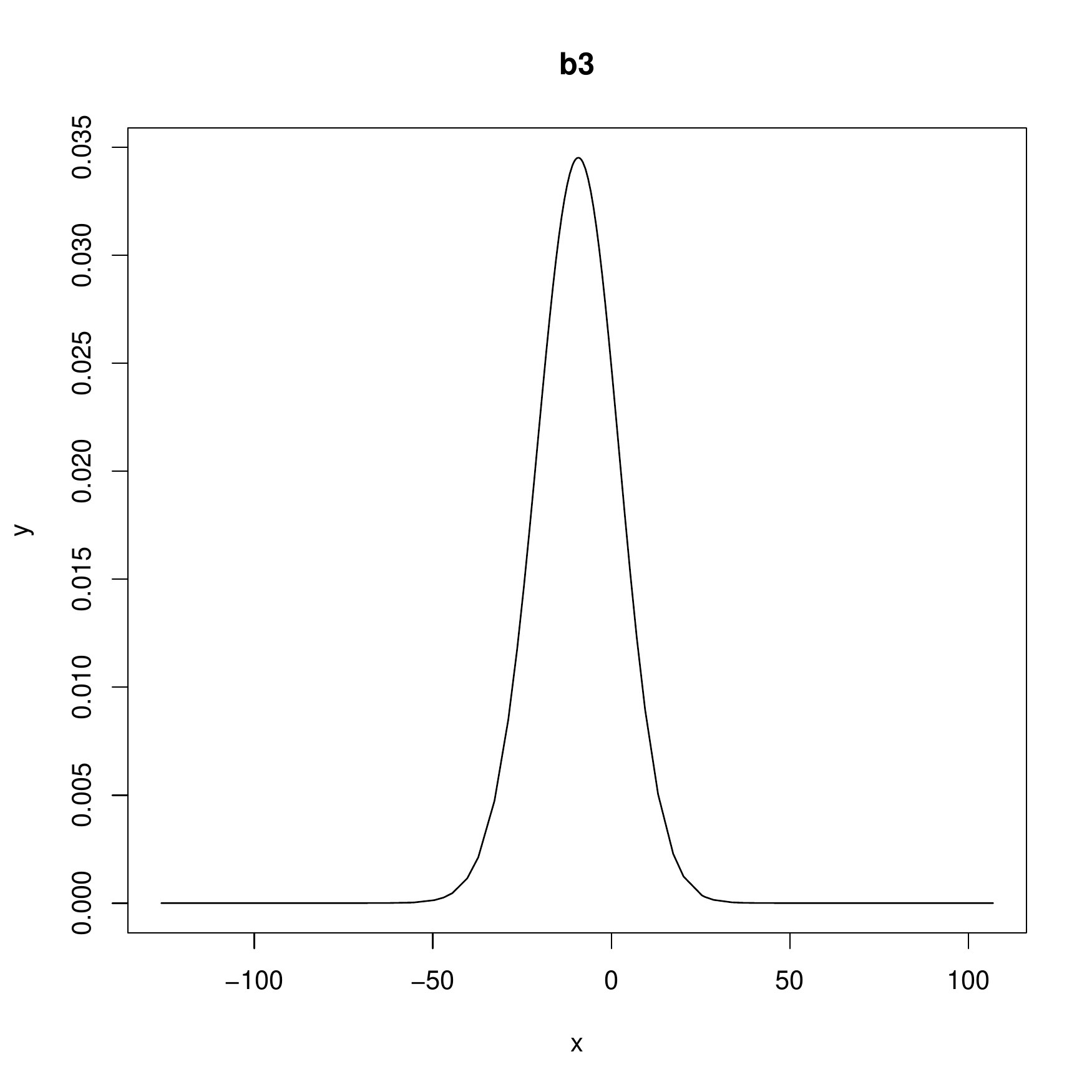}}
		\subfloat[$r$]{\includegraphics[width=0.33333\textwidth]{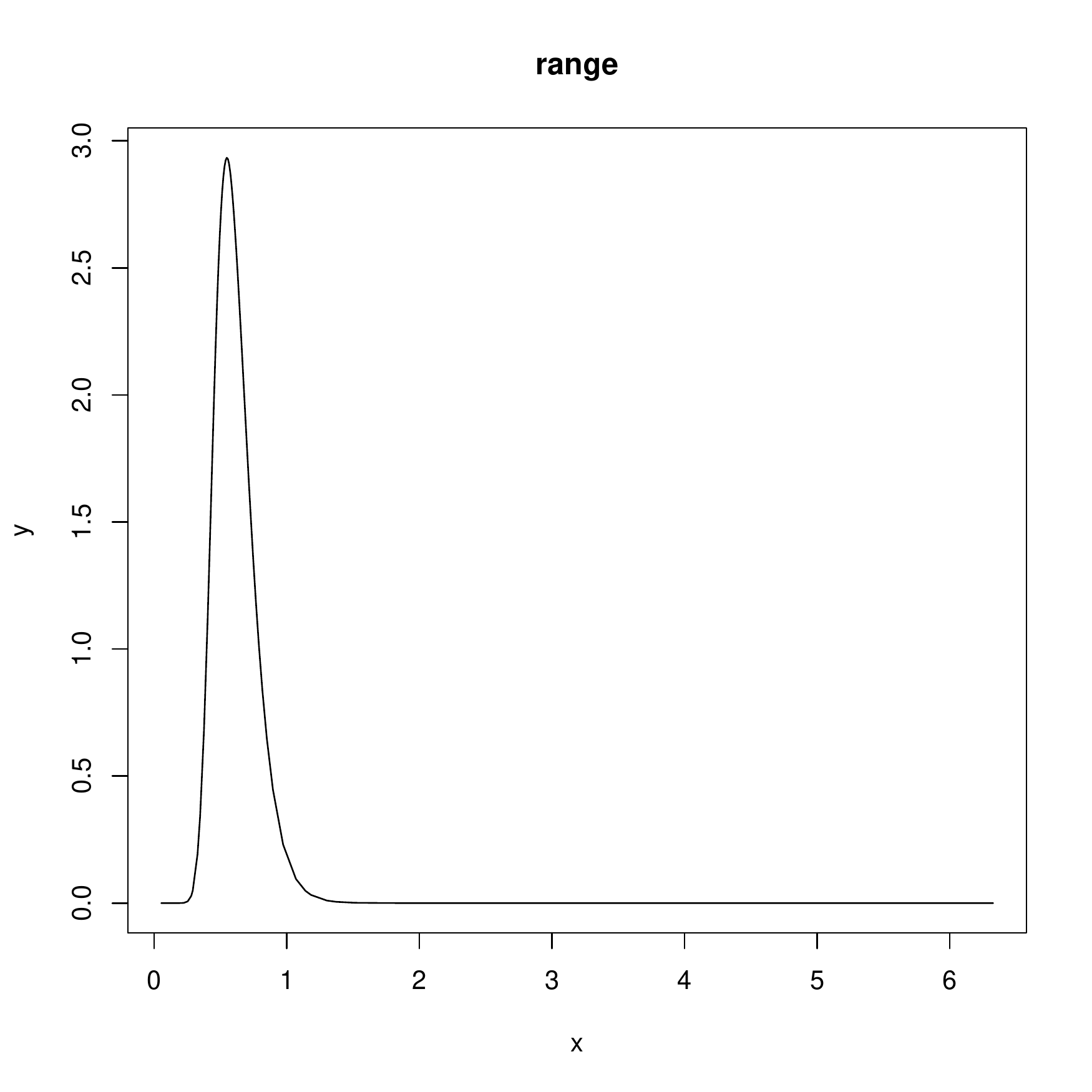}}
	\subfloat[$\sigma^2$]{\includegraphics[width=0.33333\textwidth]{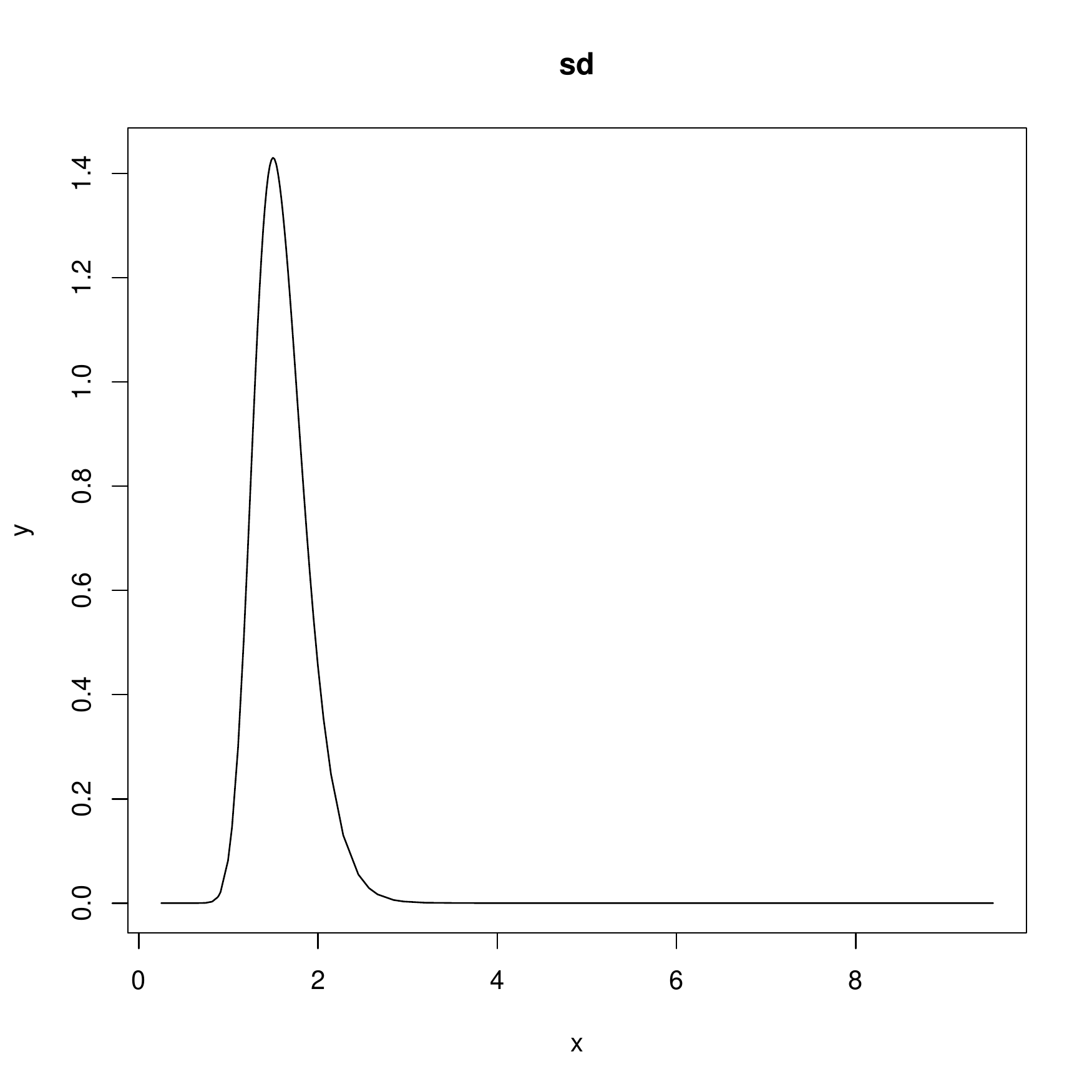}}
	\caption{Posterior distributions of the parameters and hyperparameters of the fitted LGCP model \ref{eq:selected}}
	\label{fig:plotsel}
\end{figure}

Given that, the selected model is the one in Equation (\ref{eq:selected}), where the distances from the two different sources are taken separately, together with their interaction term.
The posterior distributions of the parameters and hyperparameters of this model  are shown in Figure \ref{fig:plotsel}, and their summary statistics are reported in Table \ref{tab:tabsel1}.

\begin{table}
	\caption{Summary statistics of the posterior distributions of parameters and hyperparameters of the LGCP model \ref{eq:selected}}
		\label{tab:tabsel1}
	\begin{tabular}{rrrrrrr}
		\hline
		& mean & sd & 0.025quant & 0.5quant & 0.975quant & mode \\ 
	
		\hline
				$\beta_0$ & 2.60 & 1.09 & 0.23 & 2.66 & 4.58 & 2.79  \\ 
		$\beta_1$ & 0.09 & 3.38 & -6.52 & 0.06 & 6.81 & 0.03  \\ 
		$\beta_2$ & 5.88 & 2.85 & 0.52 & 5.79 & 11.72 & 5.62  \\ 
		$\beta_3$ & -9.56 & 11.65 & -32.76 & -9.47 & 13.06 & -9.27  \\ 
			\hline
		$r$ & 0.61 & 0.15 & 0.38 & 0.59 & 0.97 & 0.55 \\ 
		$\sigma^2$ & 1.60 & 0.30 & 1.11 & 1.57 & 2.28 & 1.50 \\ 
		\hline
	\end{tabular}

\end{table}

The posterior distributions of the hyperparameters  $r$ and $\sigma^2$, and their summary statistics, are not much different  from the ones fitted in the model \ref{eq:eq5} and the other models, not reported here for brevity.
On the basis of these results,  $\lambda(\boldsymbol{s})$ suddenly changes  over the study window, around its mean, that is the clustered structure of the point pattern is properly identified by the fitted model. 
Furthermore,  looking at the summary statistics of the posterior distributions of the parameters in Table \ref{tab:tabsel1}, we notice that the negative sign of the interaction term parameter $\beta_3$ suggests, as expected, that  moving away from a seismic source the probability of the occurrence of an earthquake decreases. 
These results are coherent with those obtained fitting a LGCP model with the same inhomogeneous intensity, through the local Palm likelihood maximization in \cite{dangelo2020local}, for the same seismic sequence. %Nevertheless, in the latter approach, as the parameters are allowed to vary spatially, it is also possible to identify the most inhomogeneous areas. Indeed,  we get a bivariate distribution of the parameters, defined on the spatial coordinates.

\subsection{Northern Italy Covid-19 disease mapping by the BYM model}

The dataset analysed in this section refers to the number of people infected by the Covid-19 in the 47  Northern Italian  provinces from  February 24th  to April 26th, 2020. The aim of this application  is to investigate the incidence of the cases, i.e. the infected people per district in this  area, that is also the most affected one in Italy and it is considered as one of the probable hot-spots of the spread of Covid-19 in Europe. A first exploratory analysis is provided splitting the time frame into two windows. Indeed,  on  the 8th of March the Italian Prime Minister announced a national lock-down action, leading people to  a sweeping  quarantine and significantly restricting the movements  of the country’s population in order to limit contagions at the epicenter of the Europe’s outbreak. The first effects of the quarantine on the virus diffusion  can be expected  after at least 14 days, and for this reason  we have chosen  March  22nd as  the change-point in the time interval. Therefore, the first time frame ranges from February 24th, that is the day when the first infected person was detected, to  March 22nd, the date from which the first effects of the lock-down were expected. We refer to this first time frame as pre-lockdown period. The second frame, that ranges from the  March 23rd to  April 26th, is referred to as post-lockdown period. 

\begin{figure}
	\centering
	\subfloat[Cases recorded during the pre-lockdown period]{\includegraphics[width=0.5\textwidth]{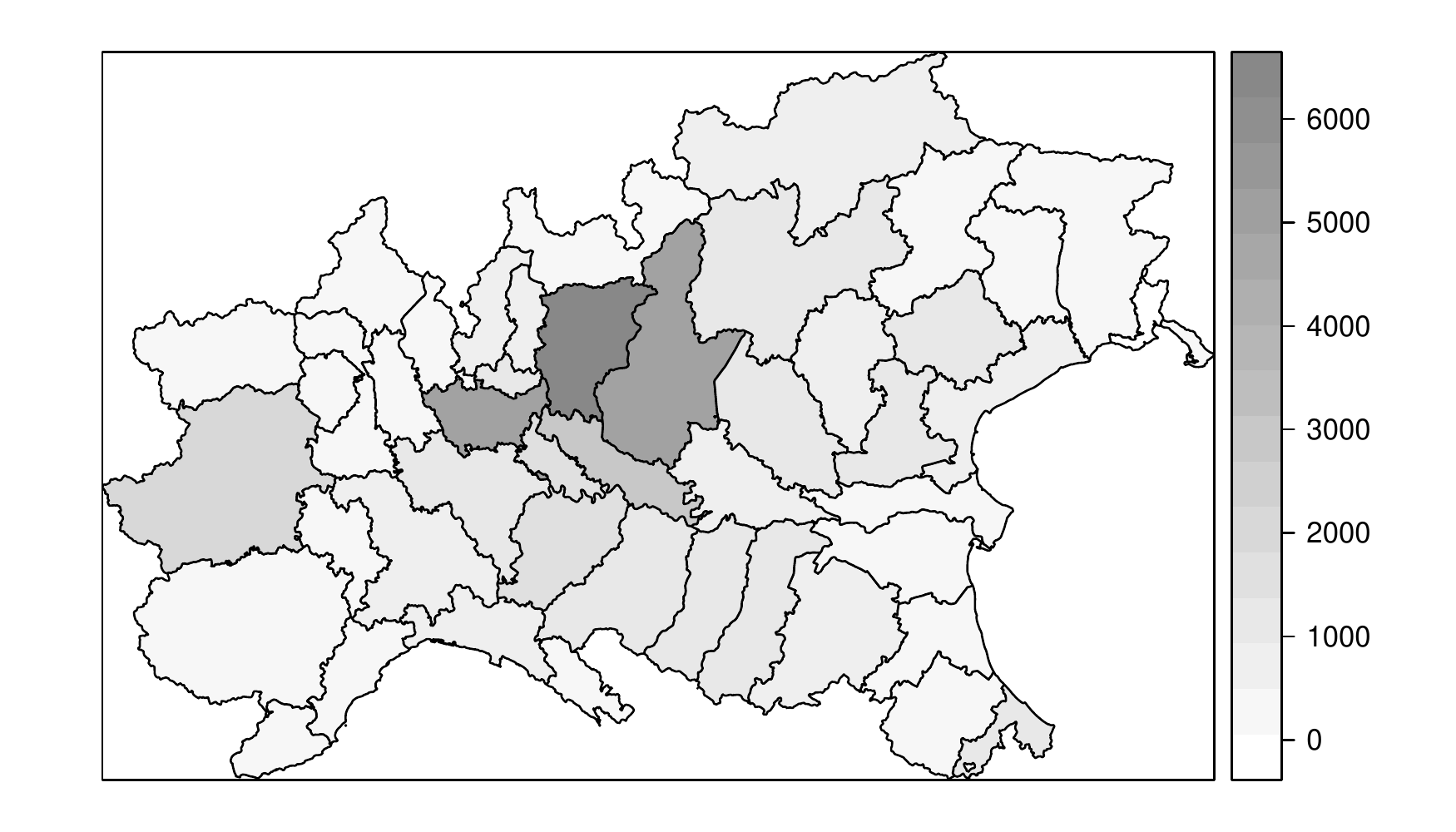}}
	\subfloat[Cases recorded during the post-lockdown period]{\includegraphics[width=0.5\textwidth]{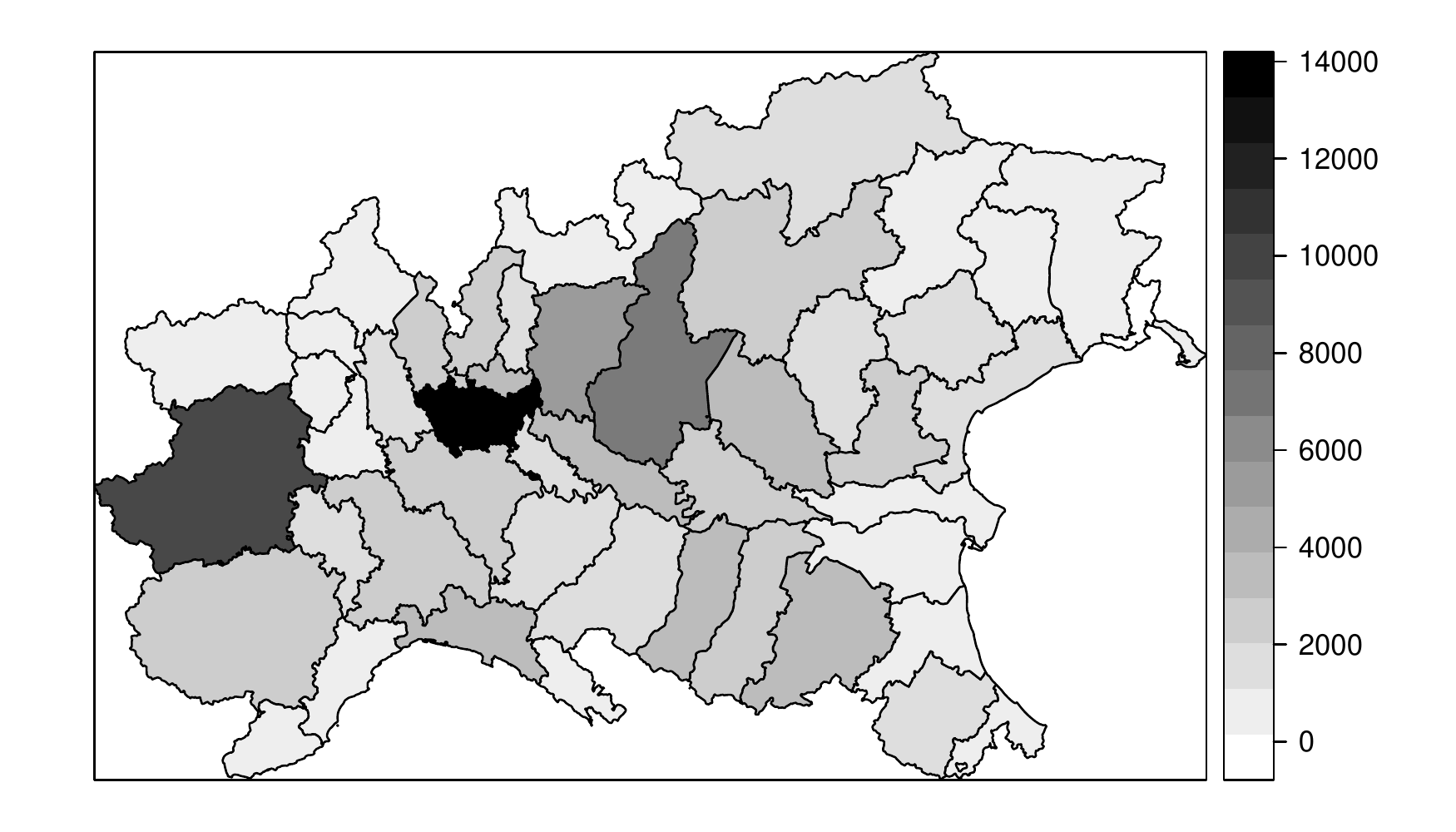}}
	\caption{Total number of cases recorded per district}
	\label{fig:covidexpl}
\end{figure}

In detail, our analysis focuses on the eight regions in the Northern Italy, namely Aosta Valley, Piedmont, Liguria, Lombardy, Emilia-Romagna, Veneto, Friuli-Venezia Giulia and Trentino-Alto Adige/Südtirol. The counts of the infected people are collected and published daily by the Italian Civil Protection at an aggregate level, both for the regions and the provinces.
In Figure \ref{fig:covidexpl}, the number of the infected people in the 47 provinces of the eight regions are shown. The darker the colour, the more the cases observed in the given district.
We know that most of the observed cases before the lock-down occurred in the district of Bergamo, in Lombardy, that is also  the darkest district in the map of Figure \ref{fig:covidexpl}(a). Bergamo is indeed a sadly well known hot-spot in the Italian spread of the Covid-19. After the lock-down, the observed cases have overall increased, and most of them are recorded in the district of Milan (Figure \ref{fig:covidexpl}(b)), meaning that most of the new infected people are detected in the closest district to Bergamo. The other most affected provinces in Lombardy are Lodi and Brescia. Outside Lombardy, the most affected district is Turin, 
capital of the Piedmont region.

In this application, two BYM models without external covariates (i.e. risk factors) are fitted, with the log-linear predictor specified as
\begin{equation}
\log \lambda_i = \alpha + v_i  +\nu_i
\label{eq:bym}
\end{equation}
to count data aggregated in the two time frames previously chosen: [February, 24th -  March, 22nd]
and [March, 23rd - April, 26th]. Therefore, we fit  two separate models referring to the two different periods,    hereafter denoted  as  pre-lockdown and post-lockdown models, respectively. According to these models,  the mean number of the cases per district, in the given time frames,
is modelled including an intercept, common to all the provinces, and  two area specific effects. These are parametrized as $u_i=v_i+\nu_i$. Furthermore, minimally informative priors are specified on the log of the unstructured and structured precisions $\log \tau_{\nu}$ and $\log \tau_v $, recalling that $\tau=\frac{1}{\sigma^2}$.
We model the number of cases per province, without external covariates, in order to focus on  the spatial variability of the phenomenon. Different analysis including  eventual risk factors could  be carried out for the regional-level data, for which further information about the cases is available.

\begin{figure}
	\centering
	\includegraphics[width=\textwidth]{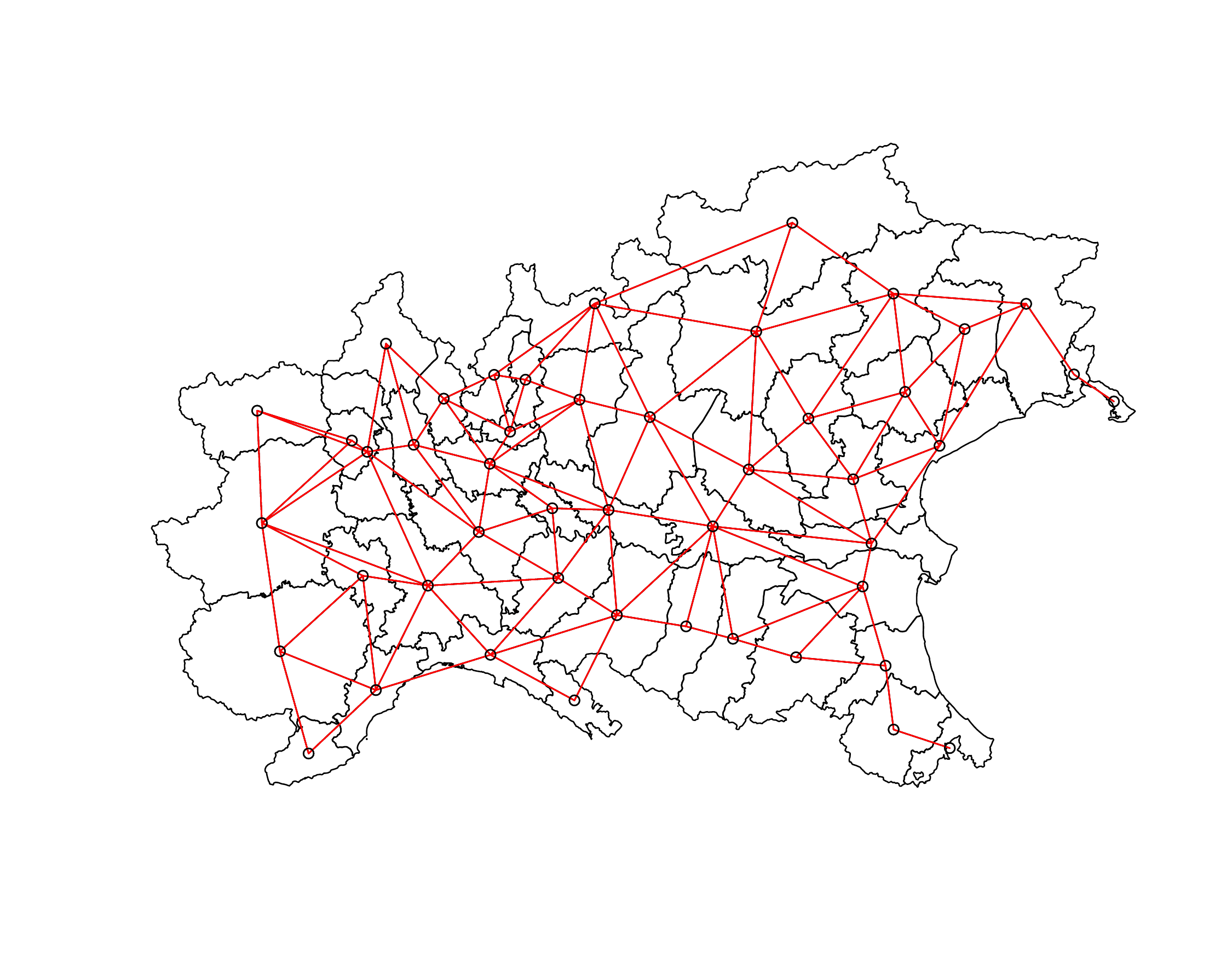}
	\caption{Northern Italy tracts linking neighbour provinces as defined in the adjacency matrix}
	\label{fig:nb}
\end{figure}

The model estimation requires the individuation of the neighbours set  for each borough, starting from the shapefile with this information on all the area boundaries. 
Therefore,  since provinces are said to be neighbours if they share a common border, and knowing that on average each district borders with 4.55 other provinces, the linking neighbour provinces are reported in Figure \ref{fig:nb}.

\begin{figure}
	\centering
	\subfloat[$\alpha$]{\includegraphics[width=0.33333\textwidth]{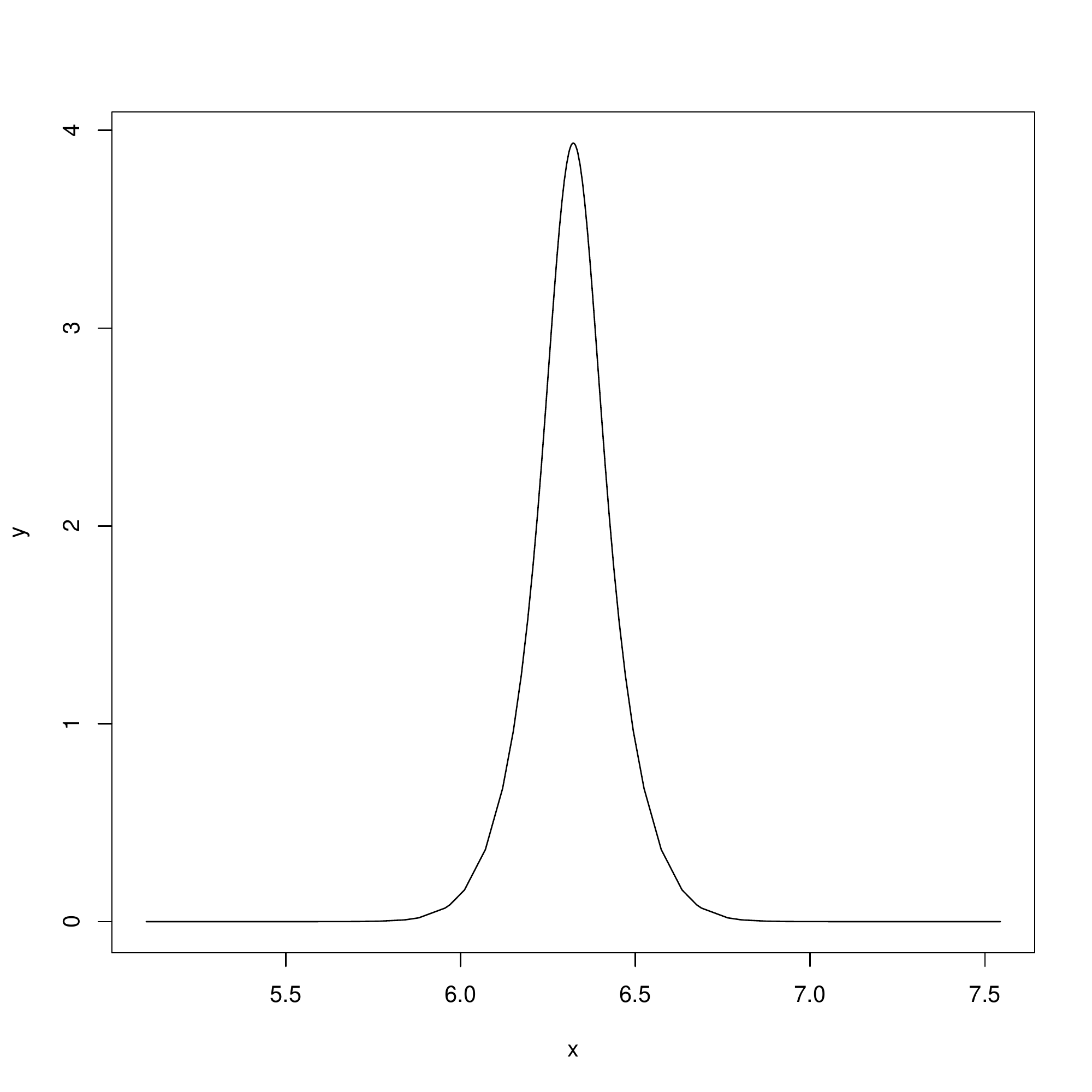}}
	\subfloat[$\sigma^2_{\nu}$]{\includegraphics[width=0.33333\textwidth]{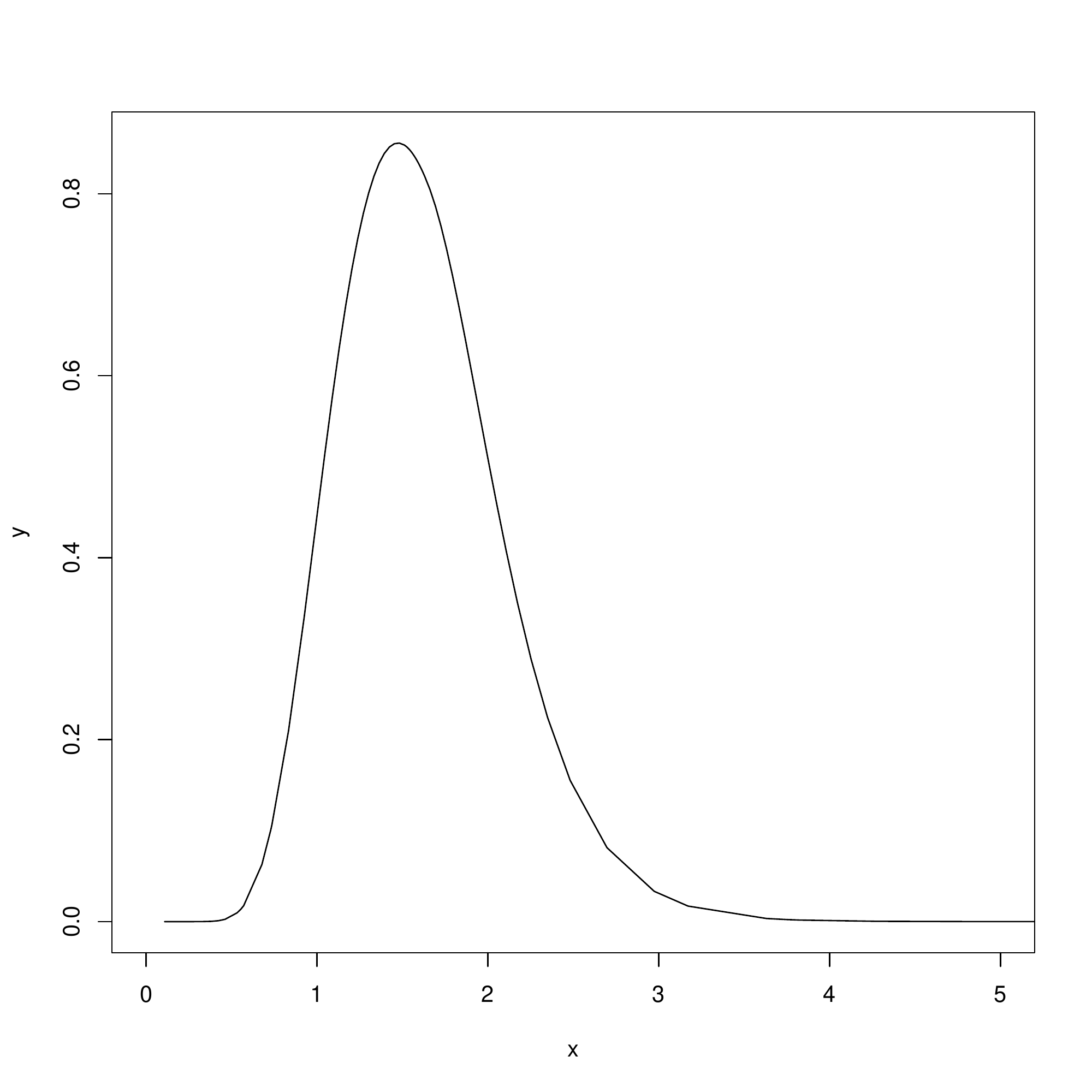}}
	\subfloat[$\sigma^2_{v}$]{\includegraphics[width=0.33333\textwidth]{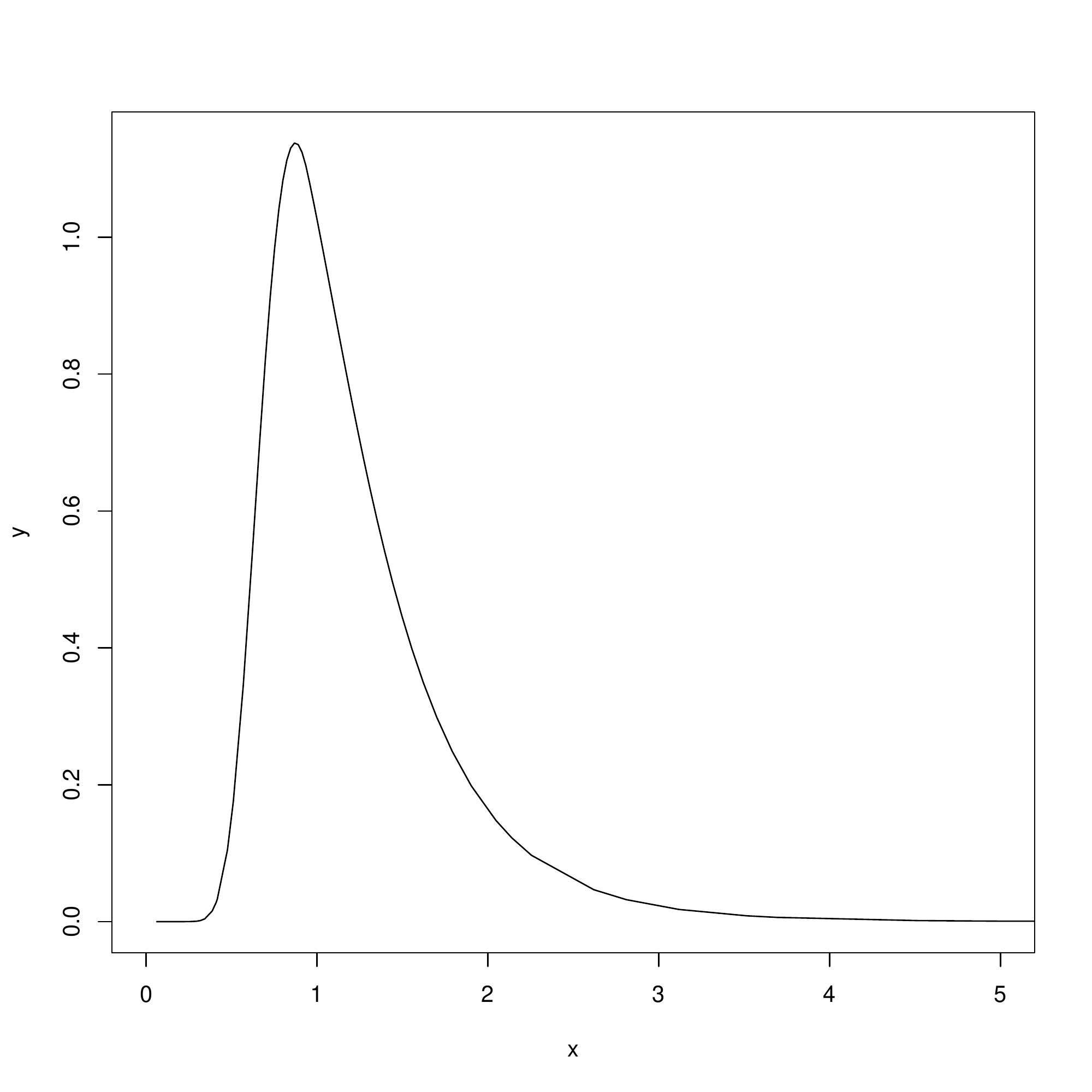}}\\
		\subfloat[$\alpha$]{\includegraphics[width=0.33333\textwidth]{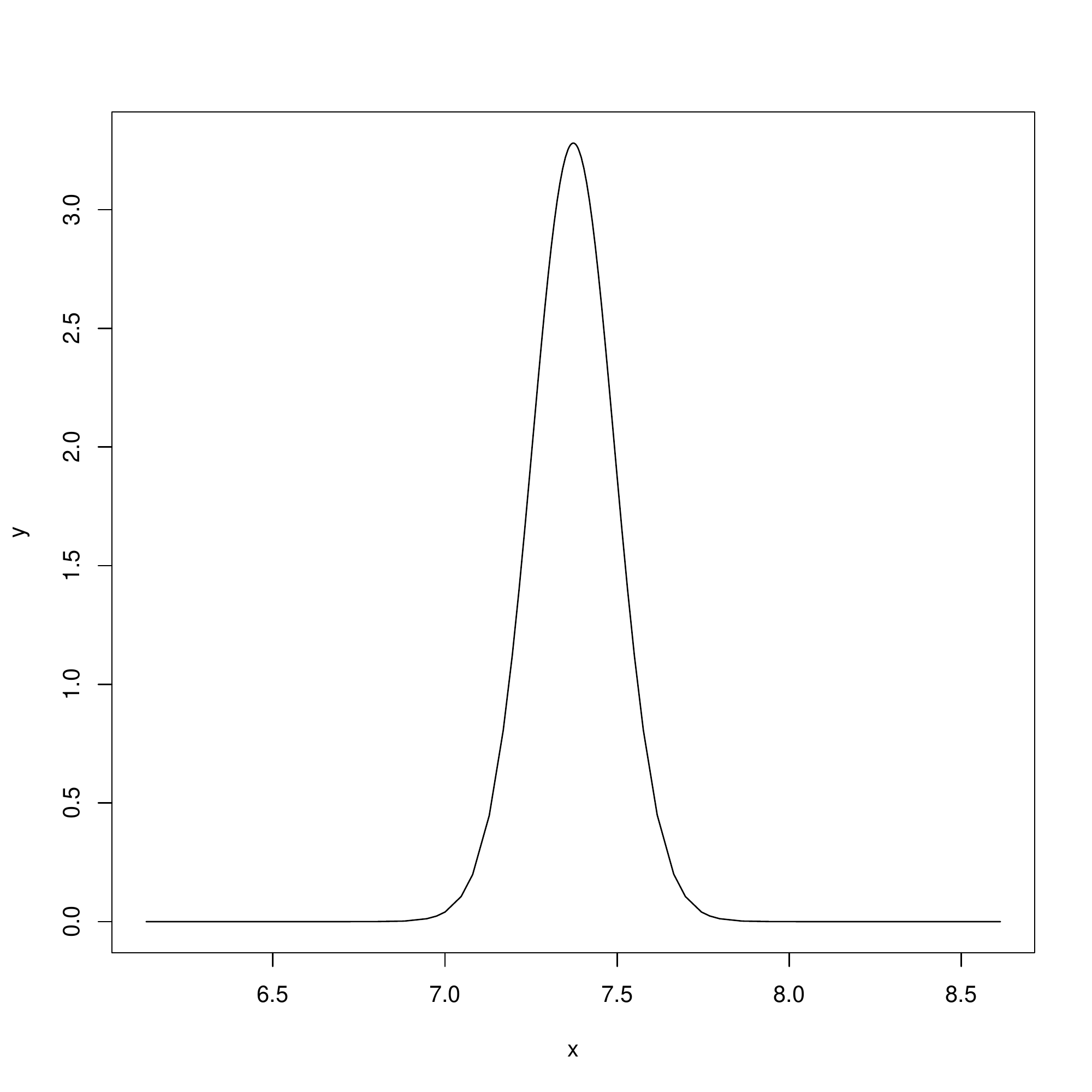}}
	\subfloat[$\sigma^2_{\nu}$]{\includegraphics[width=0.33333\textwidth]{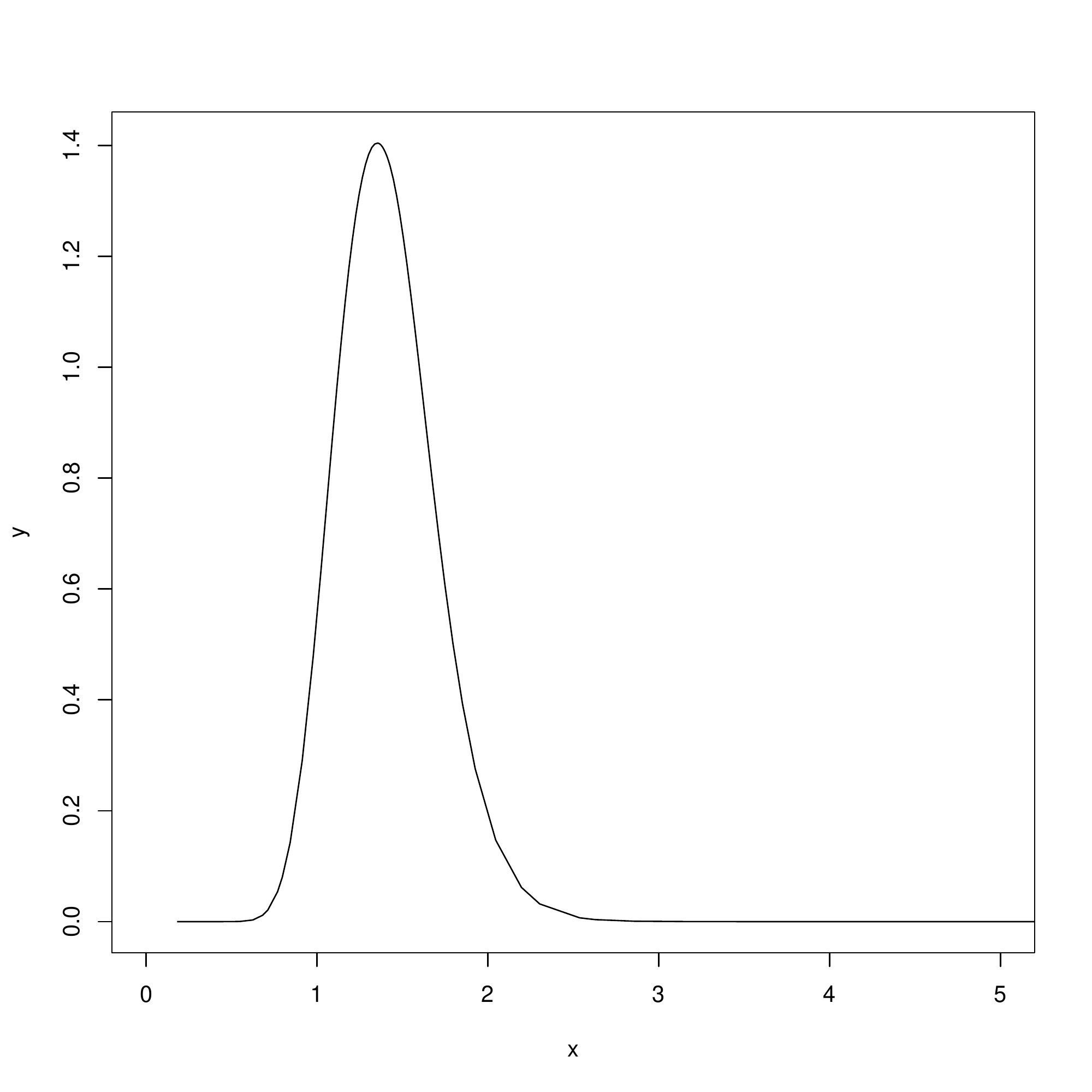}}
	\subfloat[$\sigma^2_{v}$]{\includegraphics[width=0.33333\textwidth]{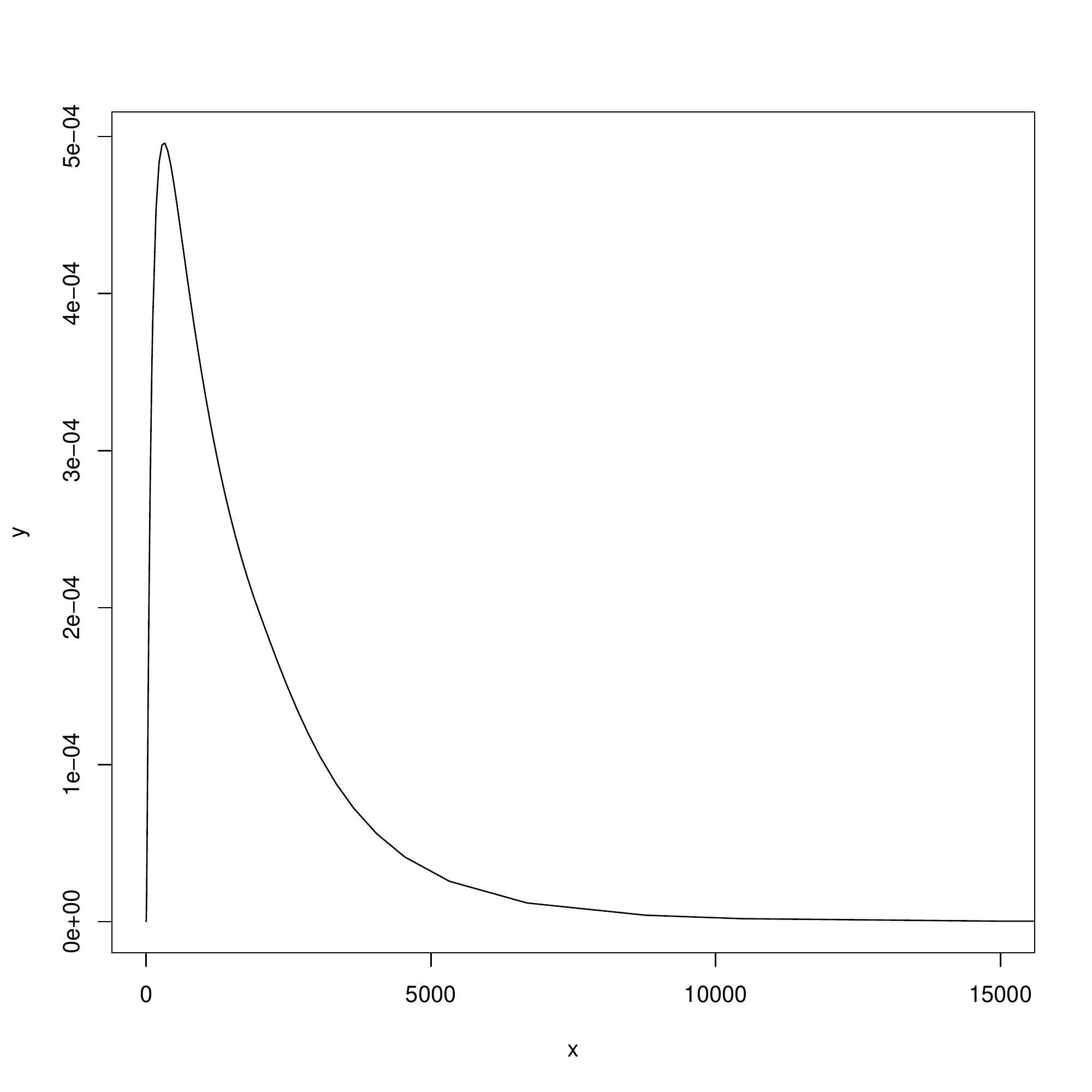}}
	\caption{Posterior distributions of the parameters and hyperparameters of the two fitted BYM models. On the top panels: the pre-lockdown BYM model ones. On the bottom panels: the post-lockdown BYM model ones.}
	\label{fig:betas1}
\end{figure}

The posterior distributions of the parameters and hyperparameters of the two models defined in Equation (\ref{eq:bym}) are shown in Figure \ref{fig:betas1}. The two hyperparameters $\sigma^2_v$ and $\sigma^2_{\nu}$ are obtained as specified in Equations (\ref{eq:str}) and (\ref{eq:unstr}), respectively. 

The summary statistics of the posterior distribution of the intercept $\alpha$, that in this context represents the expected number of cases common to all the provinces, together with the summary statistics of the posterior distributions of the hyperparameters $\sigma^2_v$ and $\sigma^2_\nu$, are reported in Tables \ref{tab:parabympre} and \ref{tab:parabympost}, for the pre- and post-lockdown models, respectively.

\begin{table}
	\caption{Summary statistics of the parameters and hyperparameters of the pre-lockdown BYM model}
		\label{tab:parabympre}
	\begin{tabular}{rrrrrrr}
		\hline
		& mean & sd & 0.025quant & 0.5quant & 0.975quant & mode \\ 
		\hline
		$\alpha$ &6.32 & 0.12 & 6.07 & 6.32 & 6.57 & 6.32\\ 
		\hline
		$\sigma^2_{\nu}$ &1.62 & 0.48 & 0.83 & 1.57 & 2.70 & 1.47 \\ 
		$\sigma^2_{v}$ & 1.22 & 0.54 & 0.57 & 1.08 & 2.62 & 0.87 \\ 
		\hline
	\end{tabular}

\end{table}

As far as for the BYM model, taking the exponential of  the mode of the posterior distribution of $\alpha$, we get the average number of counts per district, that is 561.20 with a $95\%$ credible interval ranging from 433.71 to 714.99.
The modes of the posterior distributions of the hyperparameters are  1.47 and 0.87, for the unstructured i.i.d. and the structured spatial component, respectively.

\begin{table}
\caption{Summary statistics of the parameters and hyperparameters of the  post-lockdown BYM model}
\label{tab:parabympost}
\begin{tabular}{rrrrrrr}
  \hline
 & mean & sd & 0.025quant & 0.5quant & 0.975quant & mode \\ 
  \hline
$\alpha$ &  7.37 & 0.12 & 7.13 & 7.37 & 7.62 & 7.37\\ 
 \hline
$\sigma^2_{\nu}$ &1.41 & 0.29 & 0.91 & 1.39 & 2.05 & 1.35 \\ 
$\sigma^2_{v}$ &1834.58 & 1835.99 & 116.22 & 1287.22 & 6686.58 & 311.80 \\ 
   \hline
\end{tabular}

\end{table}

Concerning the post-lockdown BYM model, the number of cases per district is 1604.27 with a $95\%$ credible interval ranging from 1248.74 to 2028.56.
In this case the modes of the posterior distributions of the hyperparameters are equal to  1.35  and 311.80, for the i.i.d. and the spatial component respectively. 

In particular, the post-lockdown model shows an higher intercept, if compared to the pre-lockdown one. This result is  reasonable,  since the number of the detected cases strictly depends on the number of taken swabs, that is largely increased from the beginning of the Italian medical emergency. Moreover, the post-lockdown model also provides an higher estimate of the mode of the posterior distribution of the precision parameter of the unstructured component, and a lower estimate of the spatial structured component, if compared to those estimated trough the pre-lockdown model.

\begin{figure}
	\centering
	\subfloat[$\zeta_i$ of the pre-lockdown BYM model]{\includegraphics[width=0.5\textwidth]{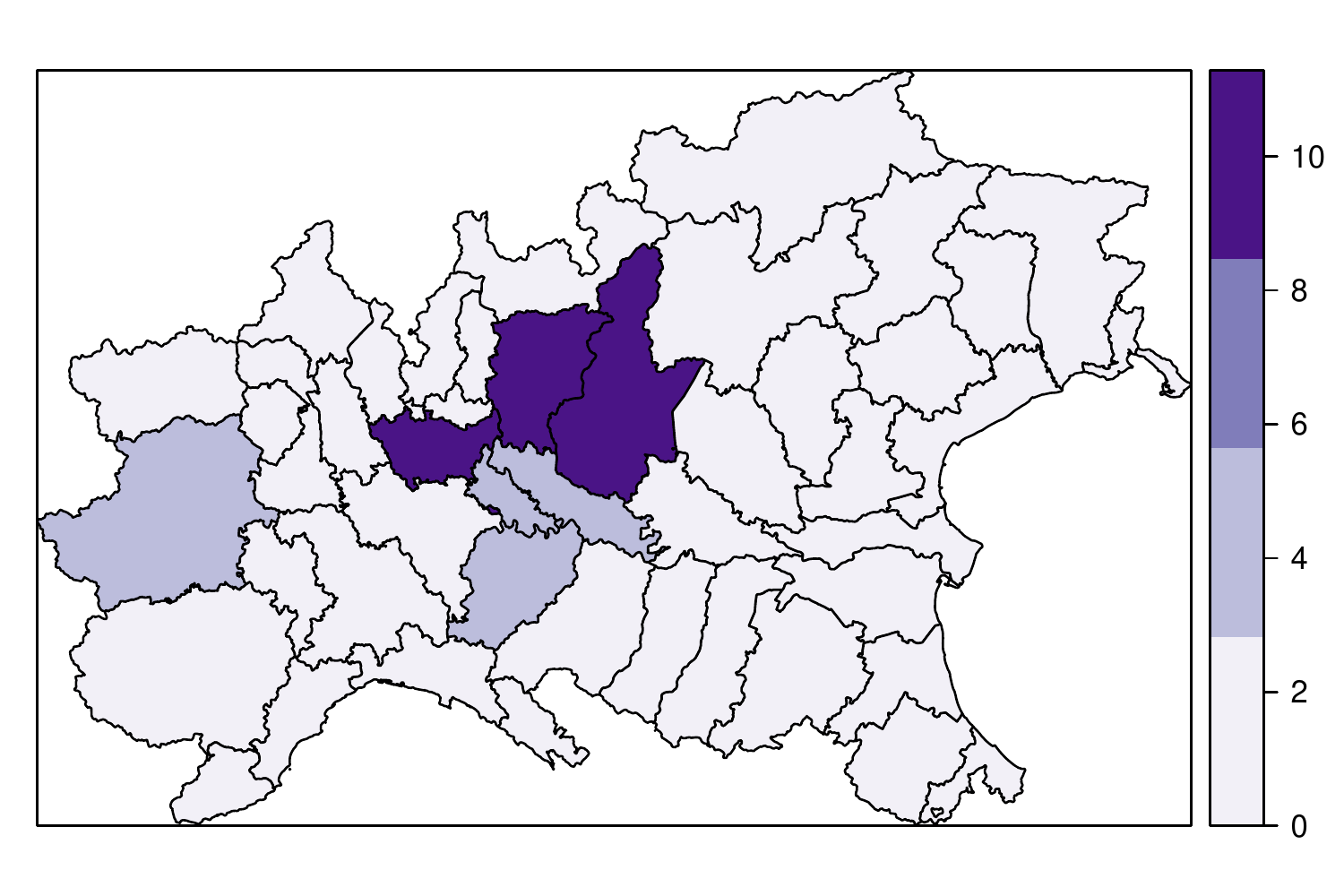}}
		\subfloat[$\zeta_i$ of the post-lockdown BYM model]{\includegraphics[width=0.5\textwidth]{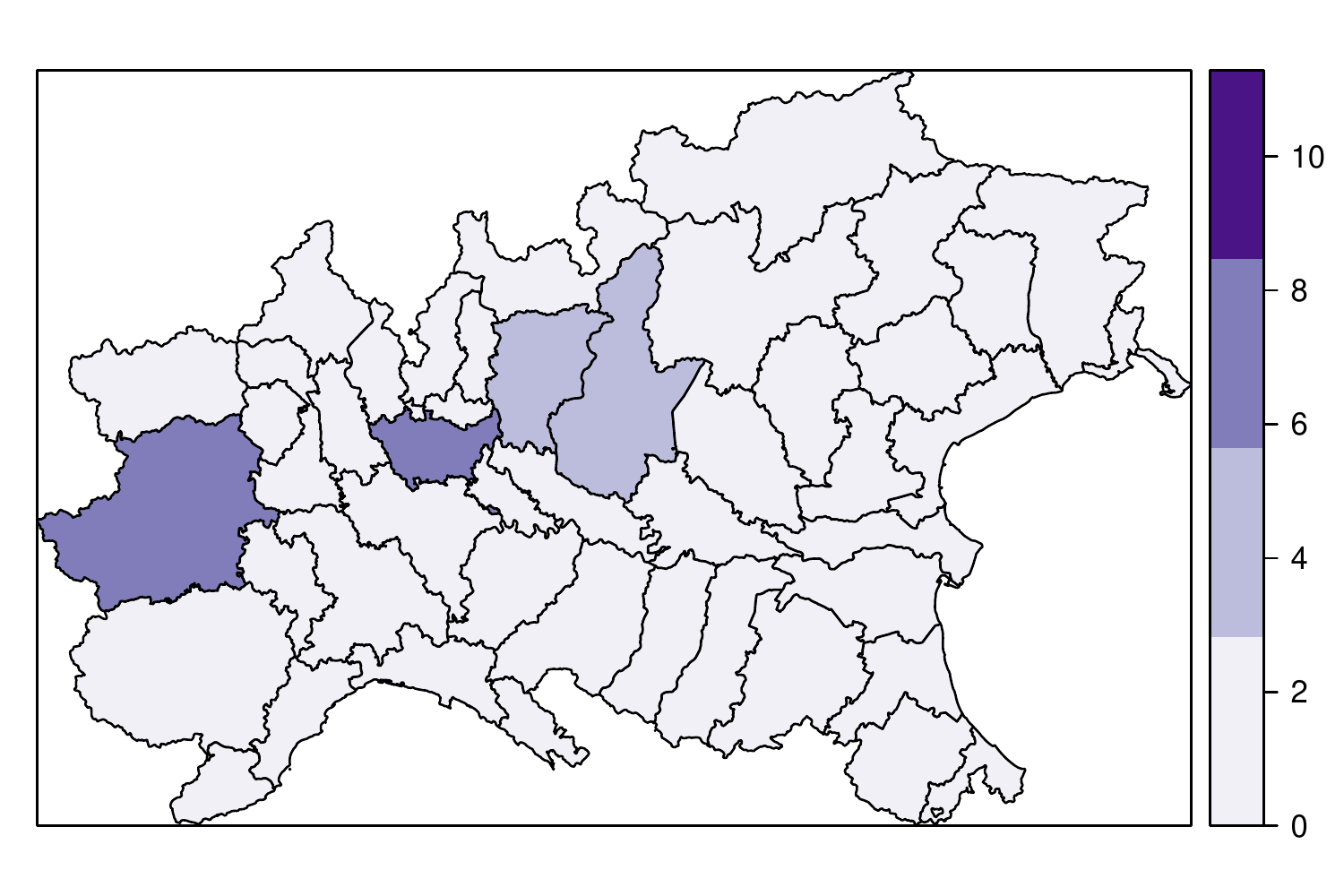}}
	\caption{District specific relative risks of cases $\zeta_i=\exp(v_i+\nu_i)$ in the disease mapping models
	}
	\label{fig:covidmods}
\end{figure}

The posterior means of the random effects $v_i$ and $\nu_i$  can be mapped, providing useful information \citep{richardson2004interpreting}.
Figure \ref{fig:covidmods} shows the map of the posterior means for the district-specific relative risk of detecting cases $\boldsymbol{\zeta}=\exp ( \boldsymbol{v}+\boldsymbol{\nu})$, compared to the whole of the  Northern Italy, of the two models. 
The darker the area, the higher the risk of infection in the given district. We may notice that in the pre-lockdown BYM model the  most risk exposed provinces are those that have experienced an higher number of cases, together with some neighbour provinces, namely Milan, Bergamo and Brescia with a relative risk higher than 8, and Turin, Lodi, Cremona and Piacenza with a relative risk higher than 5.
Moreover, although in  the post-lockdown some provinces still have an high relative risk of cases, the overall risk  is lower than the risk observed in the pre-lockdown period.

In this application the DIC values are equal to 477.49 and 526.96 for the pre-lockdown and the post-lockdown BYM models, respectively. In this particular BYM model specification, the proportion of the variance explained by
the structured spatial component can be evaluated. Since $\sigma^2_v$ and $\sigma^2_{\nu}$ are not directly comparable, we need first to get  an estimate of the posterior marginal variance for the structured effect empirically, by:
\begin{equation}
	s^2_v=\frac{\sum_{i=1}^n (v_i-\bar{v})^2}{n-1},
	\label{eq:var}
\end{equation}
where $\bar{v}$ is the average of $\boldsymbol{v}$. 
Therefore, following \cite{Blangiardo2015}, the percentage of the variance explained by the structured spatial component is obtained as the ratio between $s^2_v$  in  (\ref{eq:var}) and $s^2_v+\sigma^2_{\nu}$, with $\sigma^2_{\nu}$ the posterior marginal variance of the unstructured effect.
%\[
%\textcolor{red}{frac_{spatial}=\frac{s^2_v}{s^2_v+\sigma^2_{\nu}}.}
%\]

This percentage is 0.42 for the pre-lockdown BYM model and 0.03 for the post-lockdown one, suggesting that, while for the pre-lockdown model, a considerable part of the variability is explained by the  structured spatial component, that does not hold for the post-lockdown model.
Overall, the spatial structure  of  the pre-lockdown BYM model explains more variability than the corresponding one of the post-lockdown BYM model.

These results are coherent with the hypothesis that, after the lock-down, the spread of the Covid-19 does not seem to be mainly related to the spatial displacement of the provinces anymore, since the number of people moving among provinces has massively decreased in this second phase. However, the overall increase of the cases among the provinces may be rather caused by those district's specific issues accountable by a different class of models, operating at individual scale, rather than an aggregated one as the BYM model.

\section{Discussion and conclusion}
\label{section:conclusions}

In this paper we show some peculiarities of different complex spatial models, together with the strength of the INLA approach, by considering two different, though interesting, applications.

First, we describe a spatial seismic point pattern, focusing on a sequence of seismic events  in Greece. 
First, a LGCP model, without external covariates, is considered, where the intensity of the process only depends on the underlying spatial Gaussian Random Field. The results confirm the hypothesis that the clustered structure of the point pattern, and therefore the interaction among points, can be identified and well described by the covariance parameters.
Furthermore, accounting for the available external covariates, we show that the intensity of the given process can also depend on the distance from  external factors, as the seismic sources. In particular, the probability of the occurrence of an earthquake decreases when moving away from a seismic source. More in detail, results confirm that the two considered seismic sources  have different effects on the intensity of the process, and also their interaction  significantly characterizes the seismic generating process.
Finally, we find that the INLA-based inference, also providing the posterior distribution of parameters, yields to similar results as the ones based on the local parameter estimation proposed in \cite{dangelo2020local} for the description of the same seismic data. In that paper, the local distribution of parameters have also been provided, reflecting the typical multi-scale structure that characterizes the seismic phenomenon.
Furthermore, since combining the INLA and SPDE approaches, it would be possible to implement both spatial and spatio-temporal models for the description of point patterns, future analyses might  account for the temporal component, to provide further attempts for characterizing the complex spatio-temporal structure of the  analyzed seismic process.

To explore the broad applicability of the considered framework, we also analyse the disease count data of the Covid-19 spread out the provinces of the North of Italy, fitting a Besag-York-Mollié model. Indeed, since these infection data are available  as counts per spatial unit, the BYM-type model seems to be an obvious choice for describing them.  In particular, we show the spatial distribution of infections before and after the lock-down governmental action. Moreover, we identify the higher-risk areas by obtaining the posterior means of the district-specific relative risk of the detected cases, finding out that the  most at risk provinces are the well-known hot-spots in Lombardy, together with Turin in Piedmont. We also observe that the overall risk of infection  decreases after the lock-down, confirming the efficacy of the lockdown action taken by the Italian government.
The results confirms that, while before the lock-down the spread of the Covid-19 could be attributable to the spatial displacement of provinces, the same conclusion does not hold after the lock-down, since from that time,  the number of people moving among provinces has largely decreased. Furthermore, the BYM model allows to account for the spatial neighbourhood structure. This issue, together with the hierarchical structure  of the considered models, can be computational challenging, making the INLA approach  particularly suitable in this context.
As a future hint, the availability of geocoded health data would allow to study the epidemic phenomenon as a point pattern, with methods able to account for the self-exiting behaviour of events, as the ones reviewed in \cite{meyer2014spatio}.

Starting from  two different datasets, observed at a different levels of detail, our study aimed at showing the main advantages of using the INLA approach, that  appear to be  manly its flexibility in the presence of different complex statistical models such as complex processes in spatial  domain, both in the discrete and in the continuous domain,  and its  computational efficiency. 

\begin{thebibliography}{57}
\providecommand{\natexlab}[1]{#1}
\providecommand{\url}[1]{{#1}}
\providecommand{\urlprefix}{URL }
\expandafter\ifx\csname urlstyle\endcsname\relax
 \providecommand{\doi}[1]{DOI~\discretionary{}{}{}#1}\else
 \providecommand{\doi}{DOI~\discretionary{}{}{}\begingroup
 \urlstyle{rm}\Url}\fi
\providecommand{\eprint}[2][]{\url{#2}}

\bibitem[{Bachl et~al(2019)Bachl, Lindgren, Borchers, and
  Illian}]{bachl2019inlabru}
Bachl FE, Lindgren F, Borchers DL, Illian JB (2019) inlabru: an r package for
  bayesian spatial modelling from ecological survey data. Methods in Ecology
  and Evolution 10(6):760--766

\bibitem[{Baddeley et~al(2015)Baddeley, Rubak, and
  Turner}]{baddeley2015spatial}
Baddeley A, Rubak E, Turner R (2015) Spatial point patterns: methodology and
  applications with R. CRC Press

\bibitem[{Bakka et~al(2018)Bakka, Rue, Fuglstad, Riebler, Bolin, Illian,
  Krainski, Simpson, and Lindgren}]{bakka2018spatial}
Bakka H, Rue H, Fuglstad GA, Riebler A, Bolin D, Illian J, Krainski E, Simpson
  D, Lindgren F (2018) Spatial modeling with r-inla: A review. Wiley
  Interdisciplinary Reviews: Computational Statistics 10(6):e1443

\bibitem[{Besag et~al(1991)Besag, York, and Molli{\'e}}]{besag1991bayesian}
Besag J, York J, Molli{\'e} A (1991) Bayesian image restoration, with two
  applications in spatial statistics. Annals of the institute of statistical
  mathematics 43(1):1--20

\bibitem[{Bisanzio et~al(2011)Bisanzio, Giacobini, Bertolotti, Mosca, Balbo,
  Kitron, and Vazquez-Prokopec}]{Bisanzio2011}
Bisanzio D, Giacobini M, Bertolotti L, Mosca A, Balbo L, Kitron U,
  Vazquez-Prokopec GM (2011) {Spatio-temporal patterns of distribution of West
  Nile virus vectors in eastern Piedmont Region, Italy}. Parasites and Vectors
  4(1):230, \doi{10.1186/1756-3305-4-230},
  \urlprefix\url{http://www.parasitesandvectors.com/content/4/1/230}

\bibitem[{Blangiardo and Cameletti(2015)}]{Blangiardo2015}
Blangiardo M, Cameletti M (2015) {Spatial and Spatio-temporal Bayesian Models
  with R - INLA}. John Wiley - Sons

\bibitem[{Blangiardo et~al(2013)Blangiardo, Cameletti, Baio, and
  Rue}]{blangiardo2013spatial}
Blangiardo M, Cameletti M, Baio G, Rue H (2013) Spatial and spatio-temporal
  models with r-inla. Spatial and spatio-temporal epidemiology 4:33--49

\bibitem[{Cameletti et~al(2013)Cameletti, Lindgren, Simpson, and
  Rue}]{cameletti2013spatio}
Cameletti M, Lindgren F, Simpson D, Rue H (2013) Spatio-temporal modeling of
  particulate matter concentration through the spde approach. AStA Advances in
  Statistical Analysis 97(2):109--131

\bibitem[{Cameletti et~al(2019)Cameletti, G{\'o}mez-Rubio, and
  Blangiardo}]{cameletti2019bayesian}
Cameletti M, G{\'o}mez-Rubio V, Blangiardo M (2019) Bayesian modelling for
  spatially misaligned health and air pollution data through the inla-spde
  approach. Spatial Statistics 31:100,353

\bibitem[{Cressie(1992)}]{cressie1992statistics}
Cressie N (1992) Statistics for spatial data. Terra Nova 4(5):613--617

\bibitem[{D'Angelo et~al(2020)D'Angelo, Siino, D'Alessandro, and
  Adelfio}]{dangelo2020local}
D'Angelo N, Siino M, D'Alessandro A, Adelfio G (2020) Local spatial
  log-gaussian cox processes for seismic data. Submitted

\bibitem[{Diggle(1979)}]{diggle1979parameter}
Diggle PJ (1979) On parameter estimation and goodness-of-fit testing for
  spatial point patterns. Biometrics pp 87--101

\bibitem[{Diggle(2013)}]{diggle2013statistical}
Diggle PJ (2013) Statistical analysis of spatial and spatio-temporal point
  patterns. CRC press

\bibitem[{Diggle and Gratton(1984)}]{diggle1984monte}
Diggle PJ, Gratton RJ (1984) Monte carlo methods of inference for implicit
  statistical models. Journal of the Royal Statistical Society: Series B
  (Methodological) 46(2):193--212

\bibitem[{Eguchi(1983)}]{eguchi1983second}
Eguchi S (1983) Second order efficiency of minimum contrast estimators in a
  curved exponential family. The Annals of Statistics pp 793--803

\bibitem[{Fong et~al(2010)Fong, Rue, and Wakefield}]{Fong2010}
Fong Y, Rue H, Wakefield J (2010) {Bayesian inference for generalized linear
  mixed models}. Biostatistics 11(3):397--412,
  \doi{10.1093/biostatistics/kxp053}

\bibitem[{Freni-Sterrantino et~al(2018)Freni-Sterrantino, Ventrucci, and
  Rue}]{freni2018note}
Freni-Sterrantino A, Ventrucci M, Rue H (2018) A note on intrinsic conditional
  autoregressive models for disconnected graphs. Spatial and spatio-temporal
  epidemiology 26:25--34

\bibitem[{Fuglstad et~al(2019)Fuglstad, Simpson, Lindgren, and
  Rue}]{fuglstad2019constructing}
Fuglstad GA, Simpson D, Lindgren F, Rue H (2019) Constructing priors that
  penalize the complexity of gaussian random fields. Journal of the American
  Statistical Association 114(525):445--452

\bibitem[{G{\'o}mez-Rubio(2020)}]{gomez2020bayesian}
G{\'o}mez-Rubio V (2020) Bayesian inference with INLA. CRC Press

\bibitem[{G{\'{o}}mez-Rubio et~al(2015)G{\'{o}}mez-Rubio, Bivand, and
  Rue}]{Gomez-Rubio2015a}
G{\'{o}}mez-Rubio V, Bivand R, Rue H (2015) {A New Latent Class to Fit Spatial
  Econometrics Models with Integrated Nested Laplace Approximations}. Procedia
  Environmental Sciences 27:116--118, \doi{10.1016/j.proenv.2015.07.119},
  \urlprefix\url{http://dx.doi.org/10.1016/j.proenv.2015.07.119}

\bibitem[{G{\'o}mez-Rubio et~al(2015)G{\'o}mez-Rubio, Cameletti, and
  Finazzi}]{gomez2015analysis}
G{\'o}mez-Rubio V, Cameletti M, Finazzi F (2015) Analysis of massive marked
  point patterns with stochastic partial differential equations. Spatial
  Statistics 14:179--196

\bibitem[{Guan(2006)}]{guan2006composite}
Guan Y (2006) A composite likelihood approach in fitting spatial point process
  models. Journal of the American Statistical Association 101(476):1502--1512

\bibitem[{Guttorp and Gneiting(2006)}]{guttorp2006studies}
Guttorp P, Gneiting T (2006) Studies in the history of probability and
  statistics xlix on the matern correlation family. Biometrika 93(4):989--995

\bibitem[{Hastings(1970)}]{hastings1970monte}
Hastings WK (1970) Monte carlo sampling methods using markov chains and their
  applications. Biometrika 57(1):97--109

\bibitem[{Illian et~al(2008)Illian, Penttinen, Stoyan, and
  Stoyan}]{illian2008statistical}
Illian J, Penttinen A, Stoyan H, Stoyan D (2008) Statistical analysis and
  modelling of spatial point patterns, vol~70. John Wiley \& Sons

\bibitem[{Illian et~al(2012)Illian, S{\o}rbye, and Rue}]{illian2012toolbox}
Illian JB, S{\o}rbye SH, Rue H (2012) A toolbox for fitting complex spatial
  point process models using integrated nested laplace approximation (inla).
  The Annals of Applied Statistics pp 1499--1530

\bibitem[{Kang et~al(2020)Kang, Choi, Kim, and Choi}]{kang2020spatial}
Kang D, Choi H, Kim JH, Choi J (2020) Spatial epidemic dynamics of the covid-19
  outbreak in china. International Journal of Infectious Diseases

\bibitem[{Konstantinoudis et~al(2020)Konstantinoudis, Schuhmacher, Rue, and
  Spycher}]{konstantinoudis2020discrete}
Konstantinoudis G, Schuhmacher D, Rue H, Spycher BD (2020) Discrete versus
  continuous domain models for disease mapping. Spatial and Spatio-temporal
  Epidemiology 32:100,319

\bibitem[{Krainski et~al(2018)Krainski, G{\'o}mez-Rubio, Bakka, Lenzi,
  Castro-Camilo, Simpson, Lindgren, and Rue}]{krainski2018advanced}
Krainski ET, G{\'o}mez-Rubio V, Bakka H, Lenzi A, Castro-Camilo D, Simpson D,
  Lindgren F, Rue H (2018) Advanced spatial modeling with stochastic partial
  differential equations using R and INLA. Chapman and Hall/CRC

\bibitem[{Li et~al(2012)Li, Brown, Gesink, and Rue}]{li2012log}
Li Y, Brown P, Gesink DC, Rue H (2012) Log gaussian cox processes and spatially
  aggregated disease incidence data. Statistical methods in medical research
  21(5):479--507

\bibitem[{Liao et~al(2020)Liao, Marley, Si, Wang, Xie, Wang, and
  Tang}]{liao2020tempo}
Liao H, Marley G, Si Y, Wang Z, Xie Y, Wang C, Tang W (2020) A tempo-geographic
  analysis of global covid-19 epidemic outside of china. Available at SSRN
  3556632

\bibitem[{Lindgren et~al(2011)Lindgren, Rue, and
  Lindstr{\"o}m}]{lindgren2011explicit}
Lindgren F, Rue H, Lindstr{\"o}m J (2011) An explicit link between gaussian
  fields and gaussian markov random fields: the stochastic partial differential
  equation approach. Journal of the Royal Statistical Society: Series B
  (Statistical Methodology) 73(4):423--498

\bibitem[{Lombardo et~al(2019)Lombardo, Opitz, and
  Huser}]{lombardo2019numerical}
Lombardo L, Opitz T, Huser R (2019) Numerical recipes for landslide spatial
  prediction using r-inla: a step-by-step tutorial. In: Spatial modeling in GIS
  and R for earth and environmental sciences, Elsevier, pp 55--83

\bibitem[{Martino et~al(2011)Martino, Akerkar, and Rue}]{Martino2011}
Martino S, Akerkar R, Rue H (2011) {Approximate Bayesian Inference for Survival
  Models}. Scandinavian Journal of Statistics 38(3):514--528,
  \doi{10.1111/j.1467-9469.2010.00715.x}

\bibitem[{Martins et~al(2013)Martins, Simpson, Lindgren, and
  Rue}]{martins2013bayesian}
Martins TG, Simpson D, Lindgren F, Rue H (2013) Bayesian computing with inla:
  new features. Computational Statistics \& Data Analysis 67:68--83

\bibitem[{Meyer et~al(2014)Meyer, Held, and H{\"o}hle}]{meyer2014spatio}
Meyer S, Held L, H{\"o}hle M (2014) Spatio-temporal analysis of epidemic
  phenomena using the r package surveillance. arXiv preprint arXiv:14110416

\bibitem[{M{\o}ller(2003)}]{moller:03}
M{\o}ller J (2003) Shot noise cox processes. Advances in Applied Probability pp
  614--640

\bibitem[{M{\o}ller et~al(1998)M{\o}ller, Syversveen, and
  Waagepetersen}]{moller:98}
M{\o}ller J, Syversveen AR, Waagepetersen RP (1998) Log gaussian cox processes.
  Scandinavian Journal of Statistics 25(3):451--482

\bibitem[{Moraga(2019)}]{moraga2019geospatial}
Moraga P (2019) Geospatial health data: Modeling and visualization with R-INLA
  and shiny. CRC Press

\bibitem[{Ogata and Katsura(1991)}]{ogata1991maximum}
Ogata Y, Katsura K (1991) Maximum likelihood estimates of the fractal dimension
  for random spatial patterns. Biometrika pp 463--474

\bibitem[{Pfanzagl(1969)}]{pfanzagl1969measurability}
Pfanzagl J (1969) On the measurability and consistency of minimum contrast
  estimates. Metrika 14(1):249--272

\bibitem[{{R Core Team}(2019)}]{R}
{R Core Team} (2019) R: A Language and Environment for Statistical Computing. R
  Foundation for Statistical Computing, Vienna, Austria,
  \urlprefix\url{https://www.R-project.org/}

\bibitem[{Ram{\'\i}rez-Aldana et~al(2020)Ram{\'\i}rez-Aldana, Gomez-Verjan, and
  Bello-Chavolla}]{ramirez2020spatial}
Ram{\'\i}rez-Aldana R, Gomez-Verjan JC, Bello-Chavolla OY (2020) Spatial
  analysis of covid-19 spread in iran: Insights into geographical and
  structural transmission determinants at a province level. medRxiv

\bibitem[{Richardson et~al(2004)Richardson, Thomson, Best, and
  Elliott}]{richardson2004interpreting}
Richardson S, Thomson A, Best N, Elliott P (2004) Interpreting posterior
  relative risk estimates in disease-mapping studies. Environmental Health
  Perspectives 112(9):1016--1025

\bibitem[{Riebler et~al(2016)Riebler, S{\o}rbye, Simpson, and
  Rue}]{riebler2016intuitive}
Riebler A, S{\o}rbye SH, Simpson D, Rue H (2016) An intuitive bayesian spatial
  model for disease mapping that accounts for scaling. Statistical methods in
  medical research 25(4):1145--1165

\bibitem[{Rue et~al(2009)Rue, Martino, and Chopin}]{rue2009approximate}
Rue H, Martino S, Chopin N (2009) Approximate bayesian inference for latent
  gaussian models by using integrated nested laplace approximations. Journal of
  the royal statistical society: Series b (statistical methodology)
  71(2):319--392

\bibitem[{Schr{\"{o}}dle and Held(2011)}]{Schrodle2011}
Schr{\"{o}}dle B, Held L (2011) {Spatio-temporal disease mapping using INLA}.
  Environmetrics 22(6):725--734, \doi{10.1002/env.1065}

\bibitem[{Schr{\"{o}}dle et~al(2012)Schr{\"{o}}dle, Held, and
  Rue}]{Schrodle2012}
Schr{\"{o}}dle B, Held L, Rue H (2012) {Assessing the Impact of a Movement
  Network on the Spatiotemporal Spread of Infectious Diseases}. Biometrics
  68(3):736--744, \doi{10.1111/j.1541-0420.2011.01717.x}

\bibitem[{Siino et~al(2016)Siino, Adelfio, Mateu, Chiodi, and
  D'Alessandro}]{siino:adelfio:mateu:16}
Siino M, Adelfio G, Mateu J, Chiodi M, D'Alessandro A (2016) Spatial pattern
  analysis using hybrid models: an application to the hellenic seismicity.
  Stochastic Environmental Research and Risk Assessment 31:1633–--1648

\bibitem[{Siino et~al(2018)Siino, Adelfio, and Mateu}]{siino2018joint}
Siino M, Adelfio G, Mateu J (2018) Joint second-order parameter estimation for
  spatio-temporal log-gaussian cox processes. Stochastic environmental research
  and risk assessment 32(12):3525--3539

\bibitem[{Simpson et~al(2016)Simpson, Illian, Lindgren, S{\o}rbye, and
  Rue}]{simpson2016going}
Simpson D, Illian JB, Lindgren F, S{\o}rbye SH, Rue H (2016) Going off grid:
  Computationally efficient inference for log-gaussian cox processes.
  Biometrika 103(1):49--70

\bibitem[{Spiegelhalter et~al(2002)Spiegelhalter, Best, Carlin, and Van
  Der~Linde}]{spiegelhalter2002bayesian}
Spiegelhalter DJ, Best NG, Carlin BP, Van Der~Linde A (2002) Bayesian measures
  of model complexity and fit. Journal of the royal statistical society: Series
  b (statistical methodology) 64(4):583--639

\bibitem[{Tanaka et~al(2008)Tanaka, Ogata, and Stoyan}]{tanaka2008parameter}
Tanaka U, Ogata Y, Stoyan D (2008) Parameter estimation and model selection for
  neyman-scott point processes. Biometrical Journal: Journal of Mathematical
  Methods in Biosciences 50(1):43--57

\bibitem[{Van~Lieshout(2000)}]{van2000markov}
Van~Lieshout M (2000) Markov point processes and their applications. World
  Scientific

\bibitem[{Van~Niekerk et~al(2019)Van~Niekerk, Bakka, Rue, and
  Schenk}]{van2019new}
Van~Niekerk J, Bakka H, Rue H, Schenk L (2019) New frontiers in bayesian
  modeling using the inla package in r. arXiv preprint arXiv:190710426

\bibitem[{Wang et~al(2018)Wang, Yue, and Faraway}]{wang2018bayesian}
Wang X, Yue YR, Faraway JJ (2018) Bayesian regression modeling with INLA. CRC
  Press

\bibitem[{Zuur et~al(2017)Zuur, Ieno, and Saveliev}]{zuur2017spatial}
Zuur AF, Ieno EN, Saveliev AA (2017) Spatial, temporal and spatial-temporal
  ecological data analysis with r-inla. Highland Statistics Ltd 1

\end{thebibliography}
\end{document}